\documentclass[aps,pre,twocolumn,groupedaddress,showpacs,preprintnumbers,amsmath,
amssymb,superscriptaddress]{revtex4}

\usepackage{graphicx}
\usepackage{MnSymbol}
\def\tE{\tau_{{E}}}
\def\tD{\tau_{{D}}}
\def\TH{T_{{H}}}
\def\pF{p_{{F}}}

\newcommand{\eref}[1]{(\ref{#1})}
\newcommand{\Eref}[1]{Eq.\ (\ref{#1})}

\begin{document}

\title{Conductance fluctuations in chaotic systems with tunnel barriers}

\author{Daniel Waltner}
\author{Jack Kuipers}
\affiliation{Institut f\"ur Theoretische Physik, Universit\"at Regensburg,
D-93040 Regensburg, Germany}
\author{Philippe Jacquod}
\affiliation{Physics Department, University of Arizona,
Tucson, Arizona 85721, USA}
\affiliation{College of Optical Sciences, University of Arizona, 
Tucson, AZ 85721, USA}
\affiliation{D\'epartement de Physique Th\'eorique, Universit\'e de Gen\`eve, CH-1211 Gen\`eve}
\author{Klaus Richter}
\affiliation{Institut f\"ur Theoretische Physik, Universit\"at Regensburg,
D-93040 Regensburg, Germany}

\date{\today}

\begin{abstract}
Quantum effects are expected to disappear in the short-wavelength, semiclassical limit. As
a matter of fact, recent investigations of transport through 
quantum chaotic systems have demonstrated the exponential
suppression of the weak localization corrections to the conductance and of 
the Fano factor for shot-noise when the Ehrenfest time $\tE$ exceeds the electronic
dwell time $\tD$. On the other hand, conductance fluctuations, an effect
of quantum coherence, retain their universal value in the limit $\tE/\tD \rightarrow \infty$,
when the system is ideally coupled to external leads. Motivated by this intriguing result
we investigate conductance fluctuations through quantum chaotic cavities coupled
to external leads via (tunnel) barriers of arbitrary transparency $\Gamma$. Using the
trajectory-based semiclassical theory of transport, 
we find a linear $\tE$-dependence of the
conductance variance showing a nonmonotonous, sinusoidal behavior as a function of $\Gamma$. Most notably, we find an increase of the
conductance fluctuations with $\tE$,
above their universal value, for $\Gamma \lesssim 0.5$.
These results, confirmed by numerical simulations, show that, contrarily to the 
common wisdom, effects of quantum coherence
may increase in the semiclassical limit, under special circumstances.
\end{abstract}
\pacs{03.65.Sq, 05.45.Mt, 73.23.Ad}
\maketitle

\section{Introduction}


Since the foundation of quantum physics there has been huge interest in the
nontrivial transition from the quantum to the classical regime. An important
observation in this context is the Ehrenfest theorem stating that the dynamics
of quantum mechanical expectation values is determined by the classical
equations of motion \cite{Ehren}. Going beyond expectation values, an 
{\it Ehrenfest time} scale has been identified as the time below which the quantum time evolution is well approximated by the corresponding classical dynamics \cite{Chi}.
The Ehrenfest time is the
time it takes for the chaotic classical dynamics to stretch an initially
narrow wave packet to some relevant classical length scale such as the 
system size $L$.
Since the stretching is exponential in classically chaotic systems, 
one has
\begin{equation}
\tE=\frac{1}{\lambda}\ln\frac{\pF L}{\hbar} ,
\end{equation}
with the Lyapunov exponent $\lambda$ of the classical dynamics, the initial
spread $\hbar/\pF$ of the wave packet and the Fermi momentum $\pF$.

In recent years, there has been much interest in determining the influence
of the Ehrenfest time on stationary transport quantities such as the conductance
\cite{Aleiner,Yevtu,Ada,Bro,Jacquod,Wal1}, its variance \cite{Bro,Suk} and its behavior under
decoherence \cite{Aleiner,Tworzydlo,Altland,Cyril}, shot noise
\cite{Agam,Whi,braunetal06} and higher moments of the current \cite{Wal,Berkolaiko}, 
and on time dependent quantities such as the
spectral form factor \cite{Bro2,Wal1,Aleiner1,Tian1}, the survival probability \cite{Wal2,Gut1}
and the fidelity \cite{Gut2}. Most of these papers
used the trajectory-based semiclassical approach to transport \cite{Jal,Ric,Richter} 
which currently is the method
of choice for investigating Ehrenfest-time dependences of quantum observables \cite{Bro,Bro1}.
The leading order quantum correction 
to the conductance was found in Refs.~\cite{Aleiner,Ada,Bro,Jacquod} to decay
exponentially with the Ehrenfest time. Qualitatively speaking this can be understood
by noting first that this contribution originates from loop diagrams [see Fig.~\ref{fig:orbt}(b)],
and as such depends on the return probability, and second that, neglecting system-dependent 
nongeneric processes, this
return probability contribution essentially vanishes for times shorter than 
the Ehrenfest time. Such an
intuitive interpretation of the Ehrenfest-time dependence does not always work:
though inherently of nonclassical nature, the leading order
contribution to the conductance
variance of systems ideally coupled to external leads turns
out to be independent of the Ehrenfest time \cite{Bro,Suk}. 

\begin{figure}
\centerline{\includegraphics[width=\columnwidth]{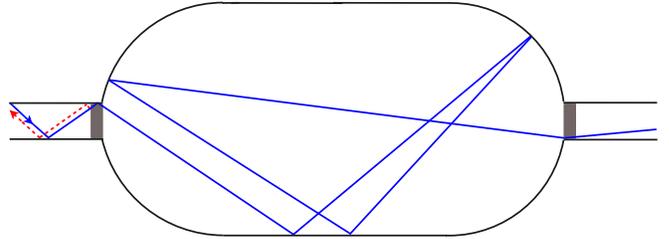}}
\caption{A chaotic cavity coupled to external leads via tunnel barriers of transparency
$\Gamma \le 1$ (gray boxes).
A classical trajectory traversing the system is shown by a solid
(blue) line. The second dashed (red) line on the left indicates a trajectory
backreflected at the barrier. This occurs with probability $1-\Gamma$.}
\label{sys}
\end{figure}

\begin{figure}
\centerline{\includegraphics[width=0.47\textwidth]{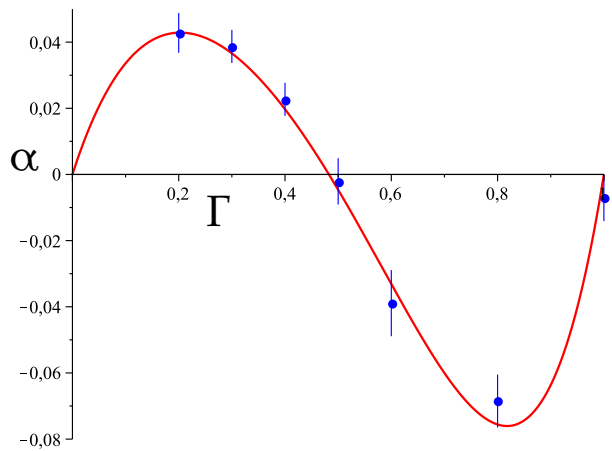}}
\caption{Dependence of the transmission fluctuations on cavity lead coupling $\Gamma$. The function ${\alpha}\tE/\tD$, defined in Eqs.\ (\ref{li22.9}) and (\ref{li23}), measures the deviation from the universal RMT variance. In this graph the parameter $\alpha$ is plotted as a function of 
the tunnel barrier transparency $\Gamma$ for time-reversal symmetric systems.
The solid red line gives the analytical prediction, Eq.\ (\ref{li23}), and the blue dots are results of
numerical simulations. Error bars indicate the standard deviation over the ensemble of
calculated data. }
\label{condfluc}
\end{figure}

In this paper, we determine semiclassically the Ehrenfest-time dependence of
the variance ${\rm var}\, G(E)$ of the conductance for a chaotic system coupled to external leads
via nonideal contacts modeled by tunnel barriers of transparency $\Gamma \le 1$.
The situation is depicted in Fig.\ \ref{sys}. The presence of tunnel barriers has the
dramatic effect that ${\rm var}\, G(E)$ increases or decreases with $\tE$, depending on 
the value of $\Gamma$. For $\Gamma \lesssim 0.5$, we even observe an enhancement of the variance
above the universal value in the presence of time reversal symmetry for equal lead widths of ${\rm var}(G)^{\rm RMT} = (1+(1-\Gamma)^2)/8$ \cite{BroRMT} 
upon increasing $\tE$. This is very surprising, given the quantal nature of the conductance 
fluctuations. In the range $0.5 \lesssim  \Gamma < 1$ we find a reduction of ${\rm var}\, G(E)$
as $\tE$ increases that is strongest around $\Gamma \simeq 0.8$, and recover the 
$\tE$-independent behavior of ${\rm var}\,G(E)$ at $\Gamma = 1$ observed in 
Refs.~\cite{Bro,Suk}. The precise dependence on $\Gamma$ is depicted in Fig.\ \ref{condfluc}.
 
To obtain these results we first, in 
Sec.\ \ref{semiapprox}, introduce the semiclassical approximation to the 
conductance variance.  Then, in Sec.\ \ref{part1}, we analytically calculate
the $\Gamma$-dependence of the diagrammatic contributions to the variance,
to leading order in the total number of open channels and linear in the Ehrenfest
time. We list all relevant diagrams and calculate their contributions. Given their 
number, we also identify the most relevant ones 
and specify the range in $\Gamma$ where they are particularly important. 
In Sec.\ \ref{part2} we present numerical results 
that confirm our analytical results and finally conclude in Sec.\ \ref{concs}.

\section{Semiclassical approach} \label{semiapprox}

Within the scattering approach to quantum transport \cite{Scat1,Scat2}, the
energy averaged conductance $G(E)$ (in units of $2e^2/h$) can be expressed in terms of the transmission
$\textbf{t}$ as
\begin{equation}
\label{in4.9}
G(E)=\left\langle{\rm
Tr}\left(\textbf{tt}^\dagger\right)\right\rangle
\end{equation}
with $\left\langle\cdots\right\rangle$ denoting an average over an energy window
that is classically small but quantum mechanically large. This 
leads to the following expression for the variance
\begin{equation}
\label{in5}
{\rm var}\,G(E)=\left\langle\left[{\rm
Tr}\left(\textbf{tt}^\dagger\right)\right]^2\right\rangle-\left\langle{\rm
Tr}\left(\textbf{tt}^\dagger\right)\right\rangle^{2}.
\end{equation}
The scattering
matrix elements are related by the Fisher-Lee relation \cite{Fisher} to the
projection of the Green function onto the transverse directions in the leads.  
Performing the projection to leading order in $\hbar$ and approximating the 
Green function semiclassically, one obtains 
\begin{equation}
\label{in6}
t_{a,b}\approx\frac{1}{\sqrt{\TH}}\sum_{\gamma(a\to b)} A_\gamma{\rm e}^{({\rm
i}/\hbar) S_\gamma} ,
\end{equation}
with the Heisenberg time $\TH$, the time conjugate to the mean level spacing.
Here the sum is over the scattering trajectories $\gamma$ which connect channel
$a$ in the entrance (or say left) lead and channel $b$ in the exit (or right)
lead in Fig.\ \ref{sys}.  The summands contain rapidly oscillating phases
depending on the classical actions $S_\gamma$ of the considered classical trajectories, and classical stability prefactors $A_\gamma$ whose
precise form is given for example in \cite{Richter}.

Inserting Eq.\ \eref{in6} into \eref{in4.9} we obtain the semiclassical
expression for the conductance,
\begin{equation}
\label{in6.9}
G(E)\approx\left\langle\frac{1}{\TH}\sum_{a,b}
\sum_{\gamma,\gamma'(a\to b)}A_{\gamma}A_{\gamma'}^{*}
{\rm e}^{\frac{{\rm i}}{\hbar}(S_{\gamma}-S_{\gamma'})}\right\rangle. 
\end{equation}
Using \eref{in6} in \eref{in5} yields the semiclassical
expression its variance,
\begin{eqnarray}
\label{in7}
{\rm var}\,G
(E)\,&\approx&\left\langle\frac{1}{\TH^{2}}
\sum_{\substack{{a,b} \cr {c,d}}}
\sum_{\substack{\gamma,\gamma' (a\to b) \cr \xi,\xi' (c\to d)}}
A_{\gamma}A_{\gamma'}^{*}A_{\xi}A_{\xi'}^{*}\right. \\
&&\hspace{3em} \times{\rm
e}^{\frac{{\rm i}}{\hbar}(S_{\gamma}-S_{\gamma'}+S_{\xi}-S_{\xi'})}
\Bigg\rangle\nonumber \\&
&-\left\langle\frac{1}{\TH}\sum_{a,b}
\sum_{\gamma,\gamma'(a\to b)}A_{\gamma}A_{\gamma'}^{*}
{\rm e}^{\frac{{\rm i}}{\hbar}(S_{\gamma}-S_{\gamma'})}\right\rangle^{2}  \nonumber
\end{eqnarray}
with the channel sums in Eqs.\ (\ref{in6.9}) and (\ref{in7}) running over all open lead channels ($N_{\rm L}$ in the left and
$N_{\rm R}$ in the right lead). If we consider contributions in the first term in Eq.\ (\ref{in7}) where $\gamma$
and $\gamma'$ form a correlated pair (with self-encounters) and $\xi$ and
$\xi'$ form a separate correlated pair, we simply recreate the second term.  We can
thus remove the second term in the above equation by removing such pairs from
the semiclassical treatment of the first term.  In terms of trajectories we then
obtain
\begin{eqnarray} \label{condvartraj}
{\rm var}\,
G(E)&\approx&\left\llangle\frac{1}{\TH^{2}}
\sum_{\substack{{a,b} \cr {c,d}}}
\sum_{\substack{\gamma,\gamma' (a\to b) \cr \xi,\xi' (c\to d)}}
A_{\gamma}A_{\gamma'}^{*}A_{\xi}A_{\xi'}^{*}\right.\nonumber\\ 
&&\hspace{3em} \times{\rm e}^{\frac{{\rm i}}{\hbar}
(S_{\gamma}-S_{\gamma'}+S_{\xi}-S_{\xi'})}\Bigg\rrangle,
\end{eqnarray}
where the trajectories $\gamma, \gamma'$ go from channel $a$ in the entrance
lead to channel $b$ in the exit lead.  Likewise trajectories $\xi, \xi'$ go from
channel $c$ to channel $d$.  Because we have removed terms from correlated
trajectories where $\gamma\approx\gamma'$ and $\xi\approx\xi'$ [this restriction
is denoted by the double bracket in Eq.\ (\ref{condvartraj})] we are left with quadruplets where all four
trajectories interact through encounters.

\begin{figure}
\centerline{\includegraphics[width=0.45\textwidth]{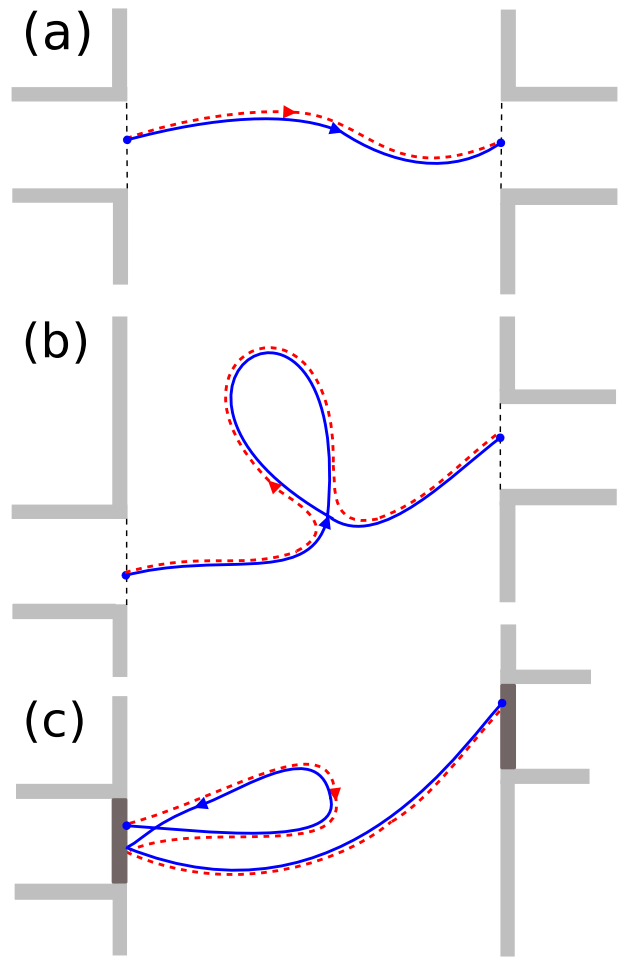}}
\caption{Schematic drawing of trajectory pairs contributing semiclassically to the energy averaged conductance (a)
A pair of identical trajectories leading to the classical contribution to the 
conductance. (b) The pair depicted here differs at a self-encounter and leads to the 
weak-localisation quantum correction to the
conductance. (c) An additional pair of orbits that needs to be considered in 
the presence of tunnel barriers.  This diagram is obtained from (b) by shrinking 
the left link until the encounter touches the tunnel barrier. An analogous 
configuration can be formed by removing the right link from Fig.\ \ref{fig:orbt}(b).}
\label{fig:orbt}
\end{figure}

Before performing the energy average the approximations for $G(E)$ and ${\rm var}\,G(E)$ in Eqs.\ (\ref{in6.9}) and (\ref{condvartraj}) are  rapidly fluctuating as a function of energy for $\hbar\rightarrow 0$. Thus only contributions from very similar trajectories survive the average. The classical contribution to Eq.\ (\ref{in6.9}) results from equal trajectories $\gamma=\gamma'$, the so called diagonal approximation \cite{Jal,Richter}, for an illustration see Fig.\ \ref{fig:orbt}(a). 
Here Eq.\ (\ref{in6.9}) yields
\begin{equation}
\label{diago}
 G(E)^{[{\ref{fig:orbt}(a)}]}=\frac{1}{\TH}\sum_{a,b}
\sum_{\gamma(a\to b)}\left|A_\gamma\right|^2.
\end{equation}
From here on, superscripts refer to the corresponding figure.
To evaluate the remaining $\gamma$-summation the sum rule for open systems \cite{Ric} that transforms the sum over orbits 
into an integral over their durations is applied 
\begin{equation}
\label{in8}
\sum_\gamma \left|A_\gamma\right|^2\approx\int_0^\infty {\rm d}t\,{\rm e}^{-t/\tD}.
\end{equation}
Here $\tD$ is the dwell time of the cavity, i.e.\ the typical time a classical particle
remains inside the chaotic system. This is given by $\tD=\TH/N$ with $N\equiv N_{\rm L}+N_{\rm R}$. This finally yields 
\begin{equation}
\label{diago1}
G(E)^{[{\ref{fig:orbt}(a)}]}\approx\frac{N_{\rm L}N_{\rm R}}{N_{\rm L}+N_{\rm R}}.
\end{equation}
Quantum corrections to this result are obtained from pairs of slightly different trajectories. The considered trajectories are almost identical differing only in how they are connected within self encounters, for the pair considered in \cite{Ric}, see Fig.\ \ref{fig:orbt}(b). Here the orbits possess close self encounters with one orbit crossing and the other anticrossing there leading to a different direction of traversal of the closed loop. Considering this pair in Eq.\ (\ref{in6.9}) leads to the leading order quantum correction to the conductance. To determine its contribution the action difference between the partner trajectories and the number of crossings needs to be determined. The calculation is done here within the phase-space approach, in the context of the conductance it was first performed in Ref.\ \cite{Heu1}. We will follow the latter approach throughout this paper. There a Poincar\'e surface of section is considered inside the encounter region and the difference along the stable and unstable directions of the piercing points of the two stretches, $\boldsymbol{s}$ and $\boldsymbol{u}$, respectively, is used to characterize an encounter. In terms of these coordinates the action difference for the orbit pair in Fig.\ \ref{fig:orbt} is given by $\Delta S=su$ \cite{Heu1}. The weight $w({s},{u})$ that additionally depends on the duration of the orbit $T$ measuring the number of encounters is obtained from the ergodicity of the flow as \cite{Heu1}
\begin{equation}
\label{weight}
w({s},{u})=\frac{\left(T-2t_{\rm enc}\right)^2}{2\Omega t_{\rm enc}},
\end{equation}
where $t_{\rm enc}\equiv 1/\lambda\ln\left(c^2/|su|\right)$ is the duration of the encounter. In general, the action difference $\Delta S$ and the weight $w(\boldsymbol{s},\boldsymbol{u})$ depend on the trajectory configuration considered. In total we obtain for the quantum correction $\delta G(E)$ resulting from the diagram in Fig.\ \ref{fig:orbt}(b)
\begin{eqnarray}
\label{quantumco}
\delta G(E)^{[{\ref{fig:orbt}(b)}]}&\approx&\int_{-c}^c\!\!{\rm d}s{\rm d}u \!\!\int_{2t_{\rm enc}}^\infty\!\!{\rm d}T \,w({s},{u}){\rm e}^{\frac{i\Delta S}{\hbar}} {\rm e}^{-\frac{\left(T-t_{\rm enc}\right)}{\tau_D}}\nonumber\\&\approx&-\frac{N_{\rm L}N_{\rm R}}{\left(N_{\rm L}+N_{\rm R}\right)^2}{\rm e}^{-\tE/\tD}
\end{eqnarray}
with $\tE\equiv1/\lambda\ln\left(c^2/\hbar\right)$. In the first line additionally the survival probability correction during the encounter \cite{Heu1} is taken into account. During the encounter the stretches are so close that the orbit escapes either during the first stretch or does not escape at all leading to the enhanced survival probability in Eq.\ (\ref{quantumco}). The ${s},{u}$-integrals in Eq.\ (\ref{quantumco}) are performed as described in \cite{Bro,Wal1}, by substituting
$su=c^2x$ and $\sigma=c/u$. The $\sigma$-integral then 
essentially cancels the $t_{{\rm enc},}$ in the denominator and the 
$x$-integral yields (after a partial integration) the contribution 
$-1/\left(\tD\TH\right){\rm e}^{-\tE/\tD}$.
   
To treat orbits differing at several places, Ref.\ \cite{Heu1} introduces the splitting of the orbit into encounters and links. During the encounters the orbits are close to themselves but the orbit and its partner are differently connected. Due to the exponential separation of neighboring trajectories in the chaotic case these last essentially an Ehrenfest time, as will become clear from the calculation below. The stretches are connected by the links, where `links' denote the long parts of 
the trajectory where the trajectory and its partner are essentially identical (up to time 
reversal symmetry). 
With a suitable change of variables in the calculation, 
in the RMT limit $\tE\rightarrow 0$, one can treat 
different encounters as distinct and separate the semiclassical contribution 
into a product of contributions over each of the links and encounters.  The total
contribution can therefore be obtained by diagrammatic rules \cite{Heu1,Berkolaiko}.

Away from this limit, and for the Ehrenfest-time dependence, the encounters may
start to overlap and for the conductance variance the trajectories can be seen 
to meet and surround periodic orbits trapped inside the system, i.e.\ they 
have encounters with periodic orbits, and must be treated as part of a continuous
family \cite{Bro}.

When we include tunnel barriers (as in Fig.\ \ref{sys}), three main changes occur 
that were originally described in \cite{Whitney}: 
\begin{enumerate}
 \item The particles enter and leave the cavity with the probability $\Gamma$, which 
leads to a factor $\Gamma^2$ for each trajectory pair. 
 \item While without tunnel barriers every particle that hits the lead leaves
the system, now only the ratio $\Gamma$ of the particles hitting the lead leaves
the cavity.  For the links, the effective dwell time is therefore 
$\tD/\Gamma$ and the dwell time in the exponential in \eref{in8} should be replaced
by this effective dwell time.  However, if trajectory stretches are correlated, as 
they are during encounters, then the whole configuration is lost if just one stretch 
of the encounter leaves the system. This happens with a probability 
$p_n\equiv1-(1-\Gamma)^n$ for $n$ correlated stretches. The dwell time in such a 
situation is therefore replaced by $\tD/\left[1-(1-\Gamma)^n\right]$. 
 \item Additional encounter diagrams become possible; for an example which contributes 
to the energy averaged conductance $G(E)$, see Fig.\ \ref{fig:orbt}(c). In this case one encounter stretch can be moved into the lead forcing the other to be backreflected at the opening. Note that configurations where both stretches are backreflected at the opening are already taken into account by the modified dwell time explained above.
\end{enumerate}
%

Although the effective dwell times are altered by the tunnel barriers, the action 
difference and weight functions are unaffected so that in the RMT limit ($\tE\rightarrow 0$)
contributions can still be obtained by diagrammatic rules: The contribution of each link
is now given by $(\Gamma N)^{-1}$. The stretches of an
encounter of $n$ orbits yield $-p_nN$. For the Ehrenfest-time dependence, however, 
these changes render the calculation of the contributions to the conductance variance 
much more difficult compared to $\Gamma=1$: Due to the discontinuous form of the effective 
dwell time, the contributions from diagrams with a different number of surroundings 
of trapped periodic orbits need to be split and treated separately. Also the possibility 
for encounter stretches to be backreflected at the tunnel barriers increases the number 
of diagrams considerably. 

\section{Diagrammatic contributions}\label{part1}

Here we calculate the leading order in $N$ contributions to the variance of the
conductance for non-zero Ehrenfest time in the presence of tunnel barriers. We 
show all the relevant diagrams and calculate their contributions.
The results given here are valid in the unitary case; results for the orthogonal
case can be obtained by multiplying the total by a factor of $2$.  

\subsection{Discrete encounters} \label{discrete}

\begin{figure}
\centerline{\includegraphics[width=\columnwidth]{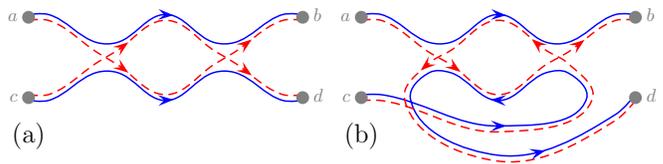}}
\caption{(a) A diagram possessing two 2-encounters in a row.  This diagram does not 
require time-reversal symmetry unlike the corresponding diagram in (b).}
\label{ana1}
\end{figure}

We start with the contributions important in the RMT limit which allow
us to recover the RMT result. The corresponding RMT calculation was performed in \cite{BroRMT} by Brouwer and Beenakker. First we consider two
2-encounters (encounters involving two trajectory stretches) in a row, see Fig.\ \ref{ana1}(a).
This diagram also occurred in the $\Gamma=1$ treatment of \cite{Bro}, but we will explain 
how for $\Gamma\neq1$ other diagrams with backreflected stretches can be derived from
this one.

When the encounters are inside the cavity and not touching the tunnel barriers, as
depicted in Fig.\ \ref{ana1}(a), we start with \Eref{condvartraj} and use the
sum rule \eref{in8}, the action difference $\Delta S(\boldsymbol{s},\boldsymbol{u})=s_1u_1+s_2u_2$
(where the subscripts refer to the different encounters) and the weight function \cite{Heu1}
\begin{equation}
\label{li0}
w(\boldsymbol{s},\boldsymbol{u})=\frac{\left(T_1-t_{{\rm enc},1}-t_{{\rm enc},2}\right)^2
\left(T_2-t_{{\rm enc},1}-t_{{\rm enc},2}\right)^2}
{4\Omega^2 t_{{\rm enc},1}t_{{\rm enc},2}},
\end{equation}
containing the durations of the two trajectories indicated by solid lines
in Fig.\ \ref{ana1}(a), $T_1$ and $T_2$, respectively and the durations of the
two encounters of the trajectories \cite{Mul1}
\begin{equation}
\label{li0.1}
t_{{\rm enc},i}\equiv\frac{1}{\lambda}\ln\frac{c^2}{|s_iu_i|},
\qquad i\in\left\lbrace1,2\right\rbrace
\end{equation}
with a classical constant $c$ again of the order one. From Eqs.\ (\ref{in7}) and (\ref{in8}) we write 
\begin{eqnarray}
\label{li0.2}
{\rm var}\,G^{[{\ref{ana1}}]}&=&
\frac{\Gamma^4N_{\rm L}^2N_{\rm R}^2}{\TH^2}\prod_{i=1}^6
\int_0^\infty {\rm d}t_i\,{\rm e}^{-\Gamma t_i/\tD}
\int_{-c}^c{\rm d}{\boldsymbol s}{\rm d}{\boldsymbol u}\frac{1}{\Omega^2}\nonumber\\
&&\times\frac{{\rm e}^{\frac{{\rm i}}{\hbar}{\boldsymbol s}{\boldsymbol u}}}
{t_{{\rm enc},1}t_{{\rm enc},2}}{\rm e}^{-p_2\left(t_{{\rm enc},1}+t_{{\rm enc},2}\right)/\tD} ,
\end{eqnarray}
with the superscript again referring to the corresponding figure: Fig.\ \ref{ana1}.
As explained in Sec.\ \ref{semiapprox}, the trajectory quadruplet leads to the
overall factor $\Gamma^4$ while the sum over possible channels provides $N_{\rm L}^2N_{\rm R}^2$.
The six links have an effective dwell time 
of $\tD/\Gamma$, while each 2-encounter experiences the dwell time $\tD/p_2$ as explained at the end of the last section.
The $s,u$-integrals are performed, as described after Eq.\ (\ref{quantumco}), by substituting
$s_iu_i=c^2x_i$ and $\sigma_i=c/u_i$. Each 
$x_i$-integral yields (after a partial integration) the contribution 
$-p_2/\left(\tD\TH\right){\rm e}^{-p_2\tE/\tD}$ as already obtained for the conductance for $\Gamma=1$. Finally the $t_i$-integrals yield
\begin{equation}
\label{li0.3}
{\rm var}\,G^{[{\ref{ana1}}]}
=\frac{N_{\rm L}^2N_{\rm R}^2}{\left(N_{\rm L}+N_{\rm R}\right)^4}
\frac{p_2^2}{\Gamma^2}{\rm e}^{-2p_2\tE/\tD},
\end{equation}
which generalizes the result for $\Gamma=1$ from \cite{Bro}.
The representation of \Eref{li0.2}, and the integrals to arrive at \eref{li0.3}, 
also nicely illustrate how the diagrammatic rules introduced above arise in this context.

We now turn to the new diagrams that arise due to the tunnel barriers where some of the 
links are shrunk until an encounter touches a barrier.  First we consider the case in
which just one link connecting the encounter to the opening is removed and the
corresponding encounter stretch now starts in the opening; as in the 
example in Fig.\ \ref{fig:orbt}(b). As the stretches during an
encounter lie very close to each other, the other encounter stretch has to be backreflected
at the opening (so only one link is lost). This contribution can therefore only exist for $\Gamma\neq1$. The
changes in the analytical calculation mainly affect the weight function, see
also \cite{Whitney,Wal1,Wal2,Gut1}.  If we shrink a link on trajectory 1 then, compared 
to \Eref{li0}, the orbit of duration $T_1$ now only involves two links so the factor
$\left(T_1-t_{{\rm enc},1}-t_{{\rm enc},2}\right)^2/2$ is replaced by
$\left(T_1-t_{{\rm enc},1}-t_{{\rm enc},2}\right)$.  Moreover, for the
encounter which touches the barrier, $t_{{\rm enc},i}$ is replaced by an integration 
variable $t'$ that is integrated from zero to the $t_{{\rm enc},i}$ defined in \eref{li0.1}.
This variable measures the length of the encounter that remains inside the system, i.e.\ which
has not yet been moved into the lead. Performing again the steps described after
\eref{li0.1} yields an expression similar to \eref{li0.2} but
with $5$ instead of $6$ link factors and $t_{{\rm enc},i}$ in the exponential in the second line
replaced by an integration variable $t'$ that is integrated from zero to
$t_{{\rm enc},i}$.  Because half of the encounter is backreflected at the tunnel barrier,
additionally this contribution is multiplied by $(1-\Gamma)$. The contribution from the 
lower limit of the $t'$-integral is zero \cite{Wal1,Wal}, leading to
\begin{equation}
\label{li0.4}
{\rm var}\,G^{[{\ref{ana1}-1{\rm l}}]}
=-\frac{4N_{\rm L}^2N_{\rm R}^2}{\left(N_{\rm L}+N_{\rm R}\right)^4}
\frac{(1-\Gamma)p_2}{\Gamma}{\rm e}^{-2p_2\tE/\tD},
\end{equation}
with the `$-1{\rm l}$' denoting that one link was removed. The prefactor 4 is due to the fact
that there are four such links we can remove.  Analogously, the
contribution where two links that connect two \textit{different} encounters to
the opening are removed is
\begin{equation}
\label{li0.5}
{\rm var}\,G^{[{\ref{ana1}-2{\rm l}}]}=
\frac{4N_{\rm L}^2N_{\rm R}^2}{\left(N_{\rm L}+N_{\rm R}\right)^4}
(1-\Gamma)^2{\rm e}^{-2p_2\tE/\tD},
\end{equation}
where there are again four possibilities of picking two links to remove.

Additionally we can, and this is a possibility which also exists for $\Gamma=1$, 
remove both links connecting the \textit{same} encounter to the opening.
This means that the trajectories tunnel straight into the encounter.  As the 
encountering orbits are so close together this means that $a=c$ or $b=d$ so that
there is only one channel summation in the lead where the encounter touches.  
Also we have two links fewer and one integral over the part of the encounter 
that remains inside the system, yielding
\begin{equation}
\label{li0.6}
{\rm var}\,G^{[{\ref{ana1}-2{\rm l(s)}}]}=
-\frac{\left(N_{\rm L}^2N_{\rm R}+N_{\rm L}N_{\rm R}^2\right)}{\left(N_{\rm L}+N_{\rm R}\right)^3}
p_2{\rm e}^{-2p_2\tE/\tD},
\end{equation}
where the additional `(s)' in the superscript indicates that two links were 
removed at the same encounter. We can further remove one link from the other 
encounter to obtain
\begin{equation}
\label{li0.65}
{\rm var}\,G^{[{\ref{ana1}-3{\rm l}}]}=
\frac{2\left(N_{\rm L}^2N_{\rm R}+N_{\rm L}N_{\rm R}^2\right)}{\left(N_{\rm L}+N_{\rm R}\right)^3}
\Gamma(1-\Gamma){\rm e}^{-2p_2\tE/\tD}.
\end{equation}
Finally, when all four links connecting the encounter to the leads are removed,
we have
\begin{equation}
\label{li0.7}
{\rm var}\,G^{[{\ref{ana1}{\rm-4l}}]}=
\frac{N_{\rm L}N_{\rm R}}{\left(N_{\rm L}+N_{\rm R}\right)^2}\Gamma^2
{\rm e}^{-2p_2\tE/\tD}.
\end{equation}

With time reversal symmetry, however, we can also have Fig.\ \ref{ana1}(b) where, 
because channels $a$ and $c$ are in the left lead and channels $b$ and $d$ in the 
right lead, we cannot shrink more than two links simultaneously. Similarly, we cannot
remove two links from the same encounter. Since $p_2=2\Gamma-\Gamma^2$,
the contributions (\ref{li0.6}--\ref{li0.7}) actually cancel so that the diagram in
Fig.\ \ref{ana1}(b) and the ones obtained by cutting links provide the same contribution as the diagram in Fig.\ \ref{ana1}(a).
Time reversal symmetry then still gives a factor 2 in this case, while for all the following 
cases, diagrams related by time reversal symmetry provide the same contributions directly.

\begin{figure}
\centerline{\includegraphics[width=\columnwidth]{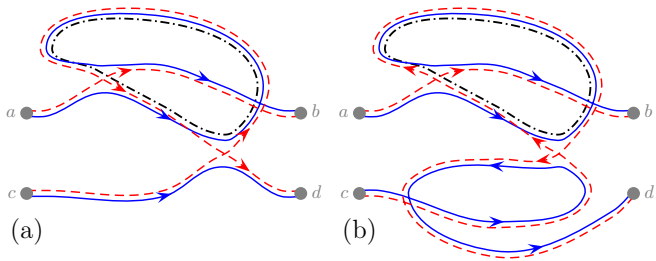}}
\caption{(a) A diagram possessing two independent 2-encounters with a
periodic orbit (dashed-dotted line). While this diagram does not require
time reversal symmetry, the corresponding diagram (b) does.}
\label{ana2}
\end{figure} 

Next we treat the diagrams with two 2-encounters which lie along a trapped periodic orbit; depicted
in Fig.\ \ref{ana2}.  Although the links connect the encounter stretches in a 
different way compared to Fig.\ \ref{ana1} these diagrams again contain two 2-encounters 
and the same number of links.   The possibilities for shrinking links are identical as for 
Fig.\ \ref{ana1}(b) and so each diagram provides the same contributions as above and the same
total contribution as the configuration in Fig.\ \ref{ana1}(a).  They are multiplied, however, 
by a factor $2$ since the orbits here have two possibilities to traverse the enclosed periodic
orbits (schematically we can also reflect the diagrams horizontally). 

\begin{figure}
\centerline{\includegraphics[width=\columnwidth]{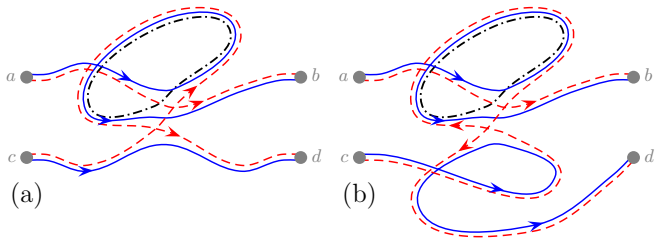}}
\caption{(a) A diagram with one 3-encounter with a periodic
orbit (dashed-dotted line).  Diagram (a) does not require time-reversal symmetry while diagram (b) does.}
\label{ana3}
\end{figure} 

The last relevant diagram type to obtain the RMT result is shown in Fig.\ \ref{ana3}.
This diagram contains one 3-encounter with a periodic orbit. Its contribution is
calculated in an analogous way to \Eref{li0.2}. Here we have one link fewer,
one encounter instead of two and involving three rather than two orbit stretches.
Therefore we have
\begin{equation}
\label{li0.8}
{\rm var}\,G^{[{\ref{ana3}}]}=
-\frac{2N_{\rm L}^2N_{\rm R}^2}{\left(N_{\rm L}+N_{\rm R}\right)^4}
\frac{p_3}{\Gamma}{\rm e}^{-p_3\tE/\tD}.
\end{equation}
Alternatively this result can be obtained from the contribution of a 3-encounter,
$K_1$, in the Appendix of \cite{Wal1} by including the correct dwell
times as well as the contributions from the links. 

Also in this case encounter stretches can be shrunk and removed: First removing one link
connecting the encounter to the opening yields
\begin{equation}
\label{li0.9}
{\rm var}\,G^{[{\ref{ana3}-1l}]}=
\frac{8N_{\rm L}^2N_{\rm R}^2}{\left(N_{\rm L}+N_{\rm R}\right)^4}
(1-\Gamma)^2{\rm e}^{-p_3\tE/\tD}
\end{equation}
and second removing both links:
\begin{equation}
\label{li0.91}
{\rm var}\,G^{[{\ref{ana3}-2l}]}=
\frac{2\left(N_{\rm L}N_{\rm R}^2+N_{\rm L}^2N_{\rm R}\right)}{\left(N_{\rm L}+N_{\rm R}\right)^3}
\Gamma(1-\Gamma){\rm e}^{-p_3\tE/\tD}.
\end{equation}

Having calculated all the contributions in the RMT limit,
we can obtain the RMT result by setting $\tE=0$.  When summing the 
the results in Eqs.\ (\ref{li0.3}--\ref{li0.91}), we obtain
\begin{eqnarray}
\label{li22.1}
{\rm var}\,G^{\rm RMT}&=&
\frac{N_{\rm L}N_{\rm R}\Gamma^6}{\left(N_{\rm L}+N_{\rm R}\right)^6}
\left[N_{\rm L}^2N_{\rm R}^2\left(4-8\Gamma+6\Gamma^2\right)\right.\nonumber\\
&&{}+\left(N_{\rm L}^3N_{\rm R}+N_{\rm L}N_{\rm R}^3\right)\left(2-2\Gamma+\Gamma^2\right)\nonumber\\
&&{}+\left.\left(N_{\rm R}^4+N_{\rm L}^4\right)\left(2\Gamma-2\Gamma^2\right)\right] ,
\end{eqnarray}
which agrees with the RMT prediction in \cite{BroRMT}.  In Appendix \ref{diffprobs},
we use these diagrams to obtain the RMT result when each channel has a different tunneling
probability.

\subsection{Periodic orbit encounters}

Having gone through all the diagrams that contribute at zero Ehrenfest time,
we now turn to those diagrams whose contribution vanishes at zero Ehrenfest time.
For these contributions, the periodic orbits in Figs.\ \ref{ana2} and \ref{ana3}
become important and we now view those diagrams as trajectories which have
an encounter with the periodic orbit, rather than with each other.  For example,
in Fig.\ \ref{ana3} we could start with the solid trajectory which passes from
channel $c$ to $d$ and the dashed trajectory from $a$ to $b$ and build the rest
of the diagram from those starting points and the periodic orbit.  Both of those 
trajectories encounter the periodic orbit once.  In the semiclassical treatment 
of Fig.\ \ref{ana3} above, it was implicitly assumed that these encounters occur
at the same point along the periodic orbit.  The resulting 3-encounter can therefore
be considered as an `aligned' 3-encounter, but for the further Ehrenfest time
dependence we also need to consider the situation where the two encounters with the 
periodic orbit occur at different points along the periodic orbit but still overlap. In this case we have a `non-aligned' 3-encounter, while when the encounters no longer
overlap we return to the two separate 2-encounters of Fig.\ \ref{ana2}.

\begin{figure}
\centerline{\includegraphics[width=0.5\columnwidth]{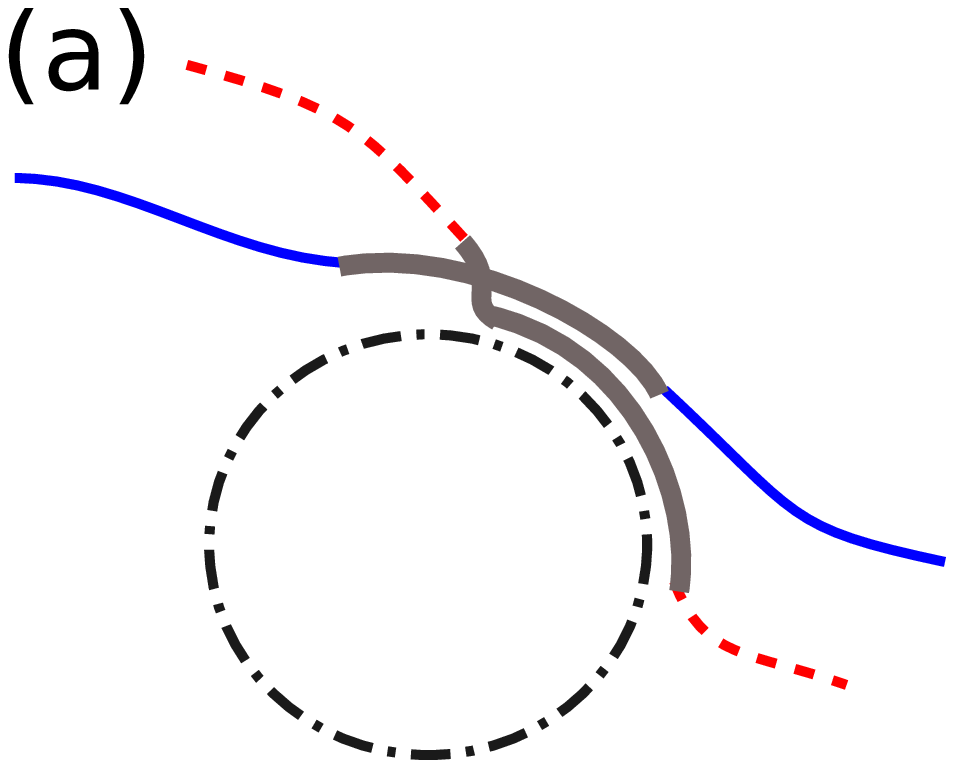}
\includegraphics[width=0.5\columnwidth]{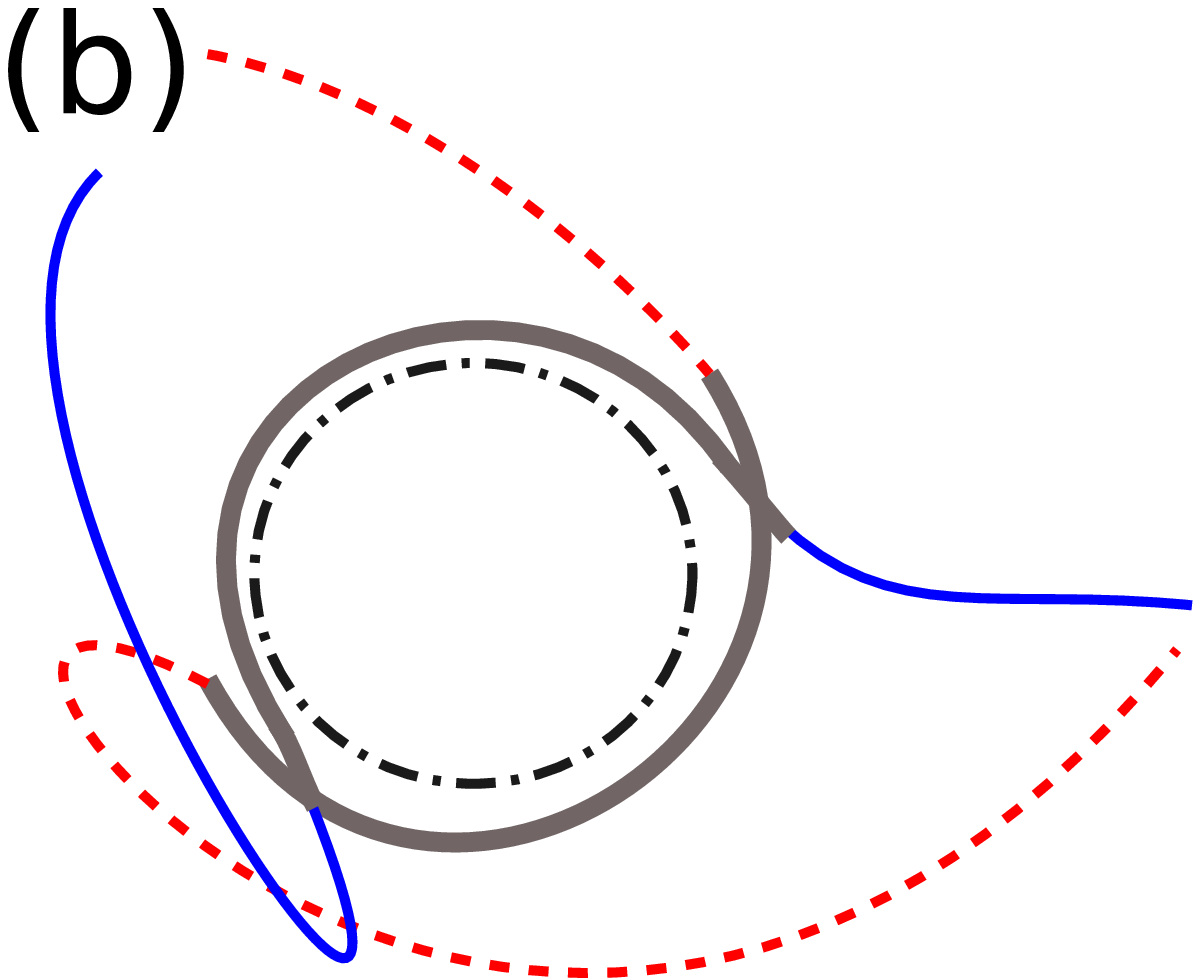}}
\caption{
Periodic orbit encounters which contribute only for nonzero
Ehrenfest time. (a) the base trajectories for a non-aligned 3-encounter and (b)
for encounters which overlap at both ends thus enclosing the periodic orbit
(depicted dashed dotted).  The encounter stretches are shown thick (gray), while
the links connecting the encounter stretches to the opening are indicated by solid 
(blue) and dashed (red) lines. To obtain the complete quadruplet of trajectories 
(with a small action difference), an additional traversal of the periodic orbit 
must be included with one of the base trajectories and included with the other base 
trajectory for the partner trajectories.}
\label{ana4}
\end{figure}

We now derive the Ehrenfest-time dependence of a non-aligned 3-encounter, whose base
trajectories are depicted in Fig.\ \ref{ana4}(a).  In order to obtain the complete trajectory 
quadruplet that contributes in \eref{condvartraj} we first need to include an
additional traversal of the periodic orbit with one of the base trajectories to obtain
the original trajectory pair [which has positive action in \eref{condvartraj}].  The partner 
pair (with negative action) is then created by including the extra
traversal of the periodic orbit with the other base trajectory.  In this way we
recover a diagram like Fig.\ \ref{ana3}(a) from Fig.\ \ref{ana4}(a) and a small action
difference (the action of the periodic orbit itself cancels).
After writing the contribution of a non-aligned 3-encounter in an analogous way as for
the aligned 3-encounter, as explained in Eqs.\ (\ref{li0}--\ref{li0.2}),
as product of link and encounter contributions, it can be evaluated 
by making use of the results for $K_2$ in the Appendix of \cite{Wal1}. 
Of course, with the tunnel barriers, the dwell times must be modified compared to
\cite{Wal1}: The link dwell time is modified by the factor $1/\Gamma$, while the parts
of the encounter where only one trajectory is correlated to the periodic orbit ---
these contributions are called fringes in \cite{Wal1} --- have the factor $1/p_2$
since we have two stretches close to each other.  Likewise, when both trajectories 
are correlated with the periodic orbit we have three stretches in total and the
corresponding factor $1/p_3$.  With these corrections, the contribution is
\begin{eqnarray}
\label{li1}
{\rm var}\,
G^{[{\ref{ana4}(a)}]}&=&\frac{4N_{\rm L}^2N_{\rm R}^2}{\left(N_{\rm L}+N_{\rm R}\right)^4}\frac{p_2^2}{
\Gamma\left(2p_2-p_3\right)}\nonumber\\
&&\times\left({\rm
e}^{-p_3\tE/\tD}-{\rm e}^{-2p_2\tE/\tD}\right). 
\end{eqnarray}

In order to proceed to the additional diagrams that arise from touching the 
tunnel barriers, we first reconsider this contribution in detail along the 
lines of \cite{Wal}.  As for Eqs.\ (\ref{li0}--\ref{li0.2}), we start with 
the weight function for two base trajectories encountering a periodic orbit,
as explained in \cite{Bro,Wal1}
\begin{eqnarray}
\label{li1.1}
w(\boldsymbol{s},\boldsymbol{u})&=&\int_0^{T_1-t_{{\rm enc},1}}{\rm d}t_1\int_0^{T_2-t_{{\rm
enc},2}}{\rm d}t_2\frac{1}{\Omega^2}\nonumber\\
&&\times\frac{1}{t_{{\rm enc},1}t_{{\rm enc},2}}\int {\rm d}\tau_p\int {\rm d}t'.
\end{eqnarray}
Here $t_1$ and $t_2$ are the durations of the links of the base trajectories
that connect the periodic orbit encounters to the lead, while $\tau_p$ is the
period of the periodic orbit.  The integral over the period corresponds to 
summing over all periodic orbits which can be encountered.  The encounter times are
as in \eref{li0.1}, but using the stable and unstable distances between the encounter
stretches and the periodic orbit itself.  Finally, 
$t'$ measures the time difference between the midpoints of the two encounter 
stretches in Fig.\ \ref{ana4}, and the $t'$-integral covers the different arrangements of
the stretches relative to each other.  The limits of the two last integrals in 
\eref{li1.1} are not given as they depend on the specific configuration considered below. 
The first two integrals in \eref{li1.1} can again be transformed, together with the 
integrals from the sum rule \eref{in8}, into a product of link and encounter 
contributions.  The general expression for the contributions to the variance 
from diagrams containing an enclosed periodic orbit that does not touch the lead
becomes
\begin{eqnarray}
\label{li1.2}
{\rm var}\,G^{\rm po}&=&\frac{\Gamma^4N_{\rm L}^2N_{\rm R}^2}{T_H^2}
\prod_{i=1}^4\int_0^\infty {\rm d}t_i\,{\rm e}^{-\Gamma t_i/\tD}
\int_{-c}^c{\rm d}\boldsymbol{s}{\rm d}\boldsymbol{u}
\frac{1}{\Omega^2} \nonumber\\
&&\times\frac{{\rm e}^{\frac{{\rm i}}{\hbar}\boldsymbol{su}}}{t_{{\rm enc},1}
t_{{\rm enc},2}}\int{\rm d}\tau_p\int {\rm d}t'
{\rm e}^{-P(\tau_p,t_{{\rm enc},1},t_{{\rm enc},2},t')}.\nonumber\\
\end{eqnarray}
Like the limits of the $\tau_p$- and $t'$-integrals, the function
$P$ determining the survival probability along the periodic orbit 
(and the encountering trajectory stretches)
will be specified for each contribution separately below. 

For example, for the non-aligned 3-encounter, we can arrange the different
alignments in terms of the durations of the two encounters.  We let
$t_{{\rm enc,max}}$ denote the longer encounter and $t_{{\rm enc,min}}$
the shorter.  In Fig.\ \ref{ana4}(a), we do not yet allow the encounter
stretches to surround the periodic orbit so we impose the restriction
$\tau_p>t_{{\rm enc,max}}$.  Then we can first consider the case that the 
shorter encounter lies inside the longer.  We will also refer to this case
later as a `generalized' 3-encounter.  The time difference between the 
midpoints of the encounters therefore satisfies 
$|t'|<\left(t_{{\rm enc,max}}-t_{{\rm enc,min}}\right)/2$, while the survival
probability function in \eref{li1.2} is given by
\begin{eqnarray}
\label{li1.75}
P&=&\frac{\Gamma\left(\tau_p-t_{{\rm enc,max}}\right)}{\tD}
+\frac{p_2\left(t_{{\rm enc,max}}-t_{{\rm enc,min}}\right)}{\tD} \nonumber \\
& & +\frac{p_3t_{{\rm enc,min}}}{\tD}.
\end{eqnarray}
The different terms simply correspond to the parts of the periodic orbit
which are followed by one, two and three trajectory stretches respectively.
As this survival probability is independent of $t'$, the $t'$-integral simply 
yields $\left(t_{{\rm enc,max}}-t_{{\rm enc,min}}\right)$. 
Performing the remaining integrals, we obtain
\begin{equation}
\label{li1.755}
-\frac{2N_{\rm L}^2N_{\rm R}^2}{\left(N_{\rm L}+N_{\rm R}\right)^4}\frac{p_3-2p_2}{\Gamma}
{\rm e}^{-p_3\tE/\tD}.
\end{equation}
Of course, the range of $t'$ includes 0, the case in which the encounters are
perfectly aligned.  \Eref{li1.755} therefore includes the contribution of the
3-encounter in Fig.\ \ref{ana3}(a), and for just the non-aligned 3-encounter 
we subtract \eref{li0.8} to give
\begin{equation}
\label{li1.756}
\frac{4N_{\rm L}^2N_{\rm R}^2}{\left(N_{\rm L}+N_{\rm R}\right)^4}\frac{p_2}{\Gamma}
{\rm e}^{-p_3\tE/\tD}.
\end{equation}

Now we consider the case in which the shorter encounter is no longer fully
inside the longer, but where the two encounters still overlap.  This can be further
separated according to the total length of the encounters, 
$t_{{\rm encs}}=t_{{\rm enc},1}+t_{{\rm enc},2}$, compared to the periodic
orbit.  When $\tau_p>t_{{\rm encs}}$, we have the range
\begin{equation}
\left|t'\right|\in\left[\left(t_{{\rm enc,max}}-t_{{\rm
enc,min}}\right)/2,t_{{\rm encs}}/2\right] .
\end{equation}
and the survival probability
\begin{eqnarray}
\label{li1.3}
P&=&\frac{\Gamma\tau_p}{\tD}+\frac{\left(p_2-\Gamma\right)}{\tD}
\left(\frac{t_{{\rm encs}}}{2}+\left|t'\right|\right)\nonumber\\
&&+\frac{\left(p_3-p_2\right)}{\tD}
\left(\frac{t_{{\rm encs}}}{2}-\left|t'\right|\right),
\end{eqnarray}
where the terms are now expressed as corrections due to additional
correlated stretches. Integrating the exponential depending on this survival
probability in \eref{li1.2} with respect to $t'$ yields 
\begin{eqnarray}
\label{li1.4}
&&\frac{2\tD}{p_3-2p_2+\Gamma}\left({\rm e}^{-\left[\left(p_2-\Gamma\right)
t_{{\rm encs}}+\Gamma\tau_p\right]/\tD}\right. \\
&&\left.-{\rm e}^{-\left[p_3t_{{\rm enc,min}}+p_2\left(t_{{\rm enc,max}}-t_{{\rm
enc,min}}\right)+\Gamma\left(\tau_p-t_{{\rm
enc,max}}\right)\right]/\tau_D}\right) .\nonumber
\end{eqnarray}
Performing the remaining integrals in \eref{li1.2} leads to
\begin{eqnarray}
\label{li1.6}
&&\frac{4N_{\rm L}^2N_{\rm R}^2}{\left(N_{\rm L}+N_{\rm R}\right)^4}\frac{1}{
\Gamma\left(p_3-2p_2+\Gamma\right)} \\
&&\times\left[p_2^2{\rm e}^{-2p_2\tE/\tD} - p_2\left(p_3-p_2-\Gamma\right)
{\rm e}^{-\left(p_3+\Gamma\right)\tE/\tD}\right] \nonumber
\end{eqnarray}

If the encounters are longer than the periodic orbit, 
$t_{{\rm enc,max}}<\tau_p<t_{{\rm encs}}$, but we still do not 
allow them to overlap at both ends, we have the restriction
\begin{equation}
\left|t'\right|\in\left[\left(t_{{\rm enc,max}}-t_{{\rm
enc,min}}\right)/2,\tau_p-t_{{\rm encs}}/2\right] ,
\end{equation}
while the survival probability remains as in \eref{li1.3}.  The 
$t'$-integral yields
\begin{eqnarray}
\label{li1.5}
&&\frac{2\tD}{p_3-2p_2+\Gamma}\left({\rm e}^{-\left[\left(p_3-p_2\right)
t_{{\rm encs}}+\left(2p_2-p_3\right)\tau_p\right]/\tD} \right. \\
&& \left.-{\rm e}^{-\left[p_3t_{{\rm enc,min}}+p_2
\left(t_{{\rm enc,max}}-t_{{\rm enc,min}}\right)
+\Gamma\left(\tau_p-t_{{\rm enc,max}}\right)\right]/\tD}\right), \nonumber 
\end{eqnarray}
which finally leads to
\begin{eqnarray}
\label{li1.7}
&&\frac{4N_{\rm L}^2N_{\rm R}^2}{\left(N_{\rm L}+N_{\rm R}\right)^4}\left\lbrace
\frac{1}{\left(2p_2-p_3\right)\left(p_3-2p_2+\Gamma\right)}\right.\nonumber\\
&&\times\left[\left(p_3-p_2\right)p_2 {\rm e}^{-p_3\tE/\tD}
-p_2^2{\rm e}^{-2p_2\tE/\tD}\right]\nonumber\\
&&-\frac{1}{\Gamma\left(p_3-2p_2+\Gamma\right)}
\left[\left(p_3-p_2\right)p_2{\rm e}^{-p_3\tE/\tD}\right.\nonumber\\
&&\left.\left.-p_2\left(p_3-p_2+\Gamma\right){\rm e}^{-\left(p_3+\Gamma\right)
\tE/\tD}\right] \right\rbrace .
\end{eqnarray}

As can be easily checked, the sum of the contributions of the
non-aligned 3-encounter in Eqs.\ (\ref{li1.756},\ref{li1.6},\ref{li1.7})
equals the contribution calculated directly in \eref{li1}.
The reason why we have discussed this more complicated route, is that
we can use it to easily calculate the contributions when we start to 
shrink links and allow the encounter to touch the tunnel barriers.
For example, if we remove one link, then by performing analogous
steps as explained before \eref{li0.4}, we obtain
\begin{eqnarray}
\label{li1.76}
{\rm var}\,G^{[{\ref{ana4}}({\rm a})-1{\rm l}]}&=&
\frac{8N_{\rm L}^2N_{\rm R}^2}{\left(N_{\rm L}^2+N_{\rm R}^2\right)^4}
\frac{p_2\left(1-\Gamma\right)}
{\left(2p_2-p_3\right)}\left[1+\left(1-\Gamma\right)\right] \nonumber \\
&&\times\left({\rm e}^{-2p_2\tE/\tD}-{\rm e}^{-p_3\tE/\tD}\right),  
\end{eqnarray}
where in the square brackets, the $1$ results from configurations where
the enclosed periodic orbit touches the tunnel barrier where only one
encounter stretch is correlated with the orbit, and the $\left(1-\Gamma\right)$
from configurations where the periodic orbit touches the tunnel barrier
while both encounter stretches are correlated with it.
Likewise, if two links connecting \textit{different} stretches to the lead 
are removed, we obtain the contribution
\begin{eqnarray}
\label{li1.8}
{\rm var}\,G^{[{\ref{ana4}}({\rm a})-2{\rm l}]}&=&
\frac{4N_{\rm L}^2N_{\rm R}^2}{\left(N_{\rm L}^2+N_{\rm R}^2\right)^4}
\frac{\Gamma\left(1-\Gamma \right)^2\left(2-\Gamma\right)^2}
{\left(2p_2-p_3\right)} \nonumber \\
&& \times\left({\rm e}^{-p_3\tE/\tD}-{\rm e}^{-2p_2\tE/\tD}\right).  
\end{eqnarray}

Next we turn to the configuration in Fig.\ \ref{ana4}(b) with encounter stretches
overlapping at both ends. We still have $t_{{\rm enc,max}}<\tau_p<t_{{\rm encs}}$,
but a different restriction on $t'$:
\begin{equation}
\left|t'\right|\in\left[\tau_p-t_{{\rm encs}}/2,\tau_p/2,\right] .
\end{equation}
The survival probability is again independent of $t'$, and given by
\begin{equation}
\label{li1.9}
P=\frac{\left(p_3-p_2\right)t_{{\rm encs}}}{\tD}+\frac{\left(2p_2-p_3\right)\tau_p}{\tD}
\end{equation}
so that the $t'$-integral yields $(t_{{\rm encs}}-\tau_p)$.  Performing the 
remaining integrals we obtain
\begin{eqnarray}
\label{li1.10}
{\rm var}\,G^{[{\ref{ana4}({\rm b})}]}&=&
\frac{2N_{\rm L}^2N_{\rm R}^2}{\left(N_{\rm L}+N_{\rm R}\right)^4}
\left[\frac{p_2^2}{\left(2p_2-p_3\right)^2}\right. \nonumber\\
&& \times \left({\rm e}^{-2p_2\tE/\tD}-{\rm e}^{-p_3\tE/\tD}\right) \nonumber\\
&&\left. +\frac{p_2\left(p_3-p_2\right)\tE}{\left(2p_2-p_3\right)\tD}
{\rm e}^{-p_3\tE/\tD} \right].
\end{eqnarray}
Again the encounter stretches can be brought to the lead by first removing
one link
\begin{eqnarray}
\label{li1.11}
{\rm var}\,G^{[{\ref{ana4}({\rm b})-1{\rm l}}]}&=&
\frac{8N_{\rm L}^2N_{\rm R}^2}{\left(N_{\rm L}+N_{\rm R}\right)^4}
\left[\frac{p_2\Gamma(1-\Gamma)^2}{\left(2p_2-p_3\right)^2}\right.\nonumber\\
&& \times \left({\rm e}^{-p_3\tE/\tD}-{\rm e}^{-2p_2\tE/\tD}\right) \nonumber \\
&&\left.-\frac{p_3\Gamma(1-\Gamma)^2\tE}{2\left(2p_2-p_3\right)\tD}
{\rm e}^{-p_3\tE/\tD} \right] , 
\end{eqnarray}
and second by removing two links connecting \textit{different} stretches to the
opening
\begin{eqnarray}
\label{li1.12}
{\rm var}\,G^{[{\ref{ana4}({\rm b})-2{\rm l}}]}&=&
\frac{8N_{\rm L}^2N_{\rm R}^2}{\left(N_{\rm L}+N_{\rm R}\right)^4}
\left[\frac {\Gamma^2(1-\Gamma)^4}{\left(2p_2-p_3\right)^2}\right.\nonumber\\
&&\times\left({\rm e}^{-2p_2\tE/\tD}-{\rm e}^{-p_3\tE/\tD}\right) \nonumber\\
&&\left.+\frac{\Gamma^2(1-\Gamma)^4\tE}{\left(2p_2-p_3\right)\tD}
{\rm e}^{-p_3\tE/\tD}\right] . 
\end{eqnarray}

\subsection{Touching both leads}

As the durations of the encounters are of the order of the Ehrenfest time, for
vanishing Ehrenfest time we only considered the situation where the encounter
stretches could be partially reflected from the tunnel barriers in \textit{one} lead.
However, for increasing Ehrenfest time configurations where encounter stretches 
are partially reflected from the tunnel barriers in \textit{both} leads,
i.e.\ where they touch the opening at both ends, become important.
If one of the 2-encounters in Fig.\ \ref{ana1} were to be partially reflected
at both ends then one of the links between the two encounters would need to
tunnel through the barrier and exit the system so that the rest of the diagram
could not be completed.  In Figs.\ \ref{ana2} and \ref{ana3}, however, as long as 
the trajectory stretches which follow the periodic orbit are reflected at the tunnel barriers
and remain in the system, we can allow the other links to tunnel through the barrier and exit
the system.  For the base trajectories in Fig.\ \ref{ana4}(a) this means we can allow
both links of one of the base trajectories, on either side of the \textit{same} encounter, to
shrink into the start and end lead.  We start with the configuration in Fig.\ \ref{ana2} and 
for the calculation we note that the contributions of the different orbit parts in 
\eref{condvartraj} are multiplicative \cite{Heu1}. We can therefore reconnect the orbits
in such a way that they split into parts whose contributions have previously been calculated.
First there is the remaining base trajectory, with a 2-encounter with the enclosed periodic 
orbit and two links connecting it to the opening. This contributes the factor
\begin{equation}
-\frac{N_{\rm L}N_{\rm R}}{\left(N_{\rm L}+N_{\rm R}\right)^2}{p_2}
{\rm e}^{-p_2\tE/\tD}.
\end{equation}
The rest of the diagram involves the periodic orbit itself and the encountering stretch
that tunnels through to start and end in the leads.  The contribution is calculated by summing
over all enclosed periodic orbits, using the sum rule \eref{in8}.  Remember, that when allowing the encounter to move into the lead an additional time integral occurred measuring the part of the stretch that lies still inside the system. Now there
are two time integrations representing the amount of the encounter which is cut short 
in each of the two leads. The first time integral cancels $t_{{\rm enc},i}$, while the second essentially 
yields a factor $\left[1-\exp\left(-p_2t_{{\rm enc},i}/\tD\right)\right]$. For the details of the calculation we refer to \cite{Bro1,Wal}.
This part of the diagram then contributes
\begin{equation}
\label{li1.13}
\frac{N_{\rm L}N_{\rm R}}{\left(N_{\rm L}+N_{\rm R}\right)^2}\frac{(1-\Gamma)^2}{p_2}
\left(1-{\rm e}^{-p_2\tE/\tD} \right),
\end{equation}
so that this configuration of Fig.\ \ref{ana2} altogether yields
\begin{eqnarray}
\label{li1.14}
{\rm var}\,G^{[{\ref{ana2}-2{\rm l(s)}}]}&=&
-\frac{4N_{\rm L}^2N_{\rm R}^2}{\left(N_{\rm L}+N_{\rm R}\right)^4}(1-\Gamma)^2
{\rm e}^{-p_2\tE/\tD}\nonumber\\
&&\times\left(1-{\rm e}^{-p_2\tE/\tD}\right),
\end{eqnarray}
where the `(s)' in the superscript denotes that the two links were removed
along the same base trajectory from the same encounter stretch. One factor 2 in the last equation derives again from the mirror symmetry of this configuration explained after Eq.\ (\ref{li0.7}), the other from the fact that each of the two stretches can touch the opening at both ends. 

Furthermore, one of the links of the other base trajectory may also be
shrunk so that the other encounter tunnels into the lead at one end.  
This contribution is
\begin{eqnarray}
\label{li1.15}
{\rm var}\,G^{[{\ref{ana2}-3{\rm l}}]}&=&
\frac{8N_{\rm L}^2N_{\rm R}^2}{\left(N_{\rm L}+N_{\rm R}\right)^4}\frac{\Gamma(1-\Gamma)^3}{p_2}
{\rm e}^{-p_2\tE/\tD}\nonumber\\
&&\times\left(1-{\rm e}^{-p_2\tE/\tD}\right).
\end{eqnarray}
Removing all 4 links, so that both encounter stretches on the base trajectories
tunnel into the leads at both ends, is also possible. The contribution is simply 
given by the square of \eref{li1.13}.

We can repeat this treatment for the aligned 3-encounter of Fig.\ \ref{ana3}. 
If one encounter stretch tunnels into the leads at both ends we obtain
\begin{eqnarray}
\label{li1.18} 
{\rm var}\,G^{[{\ref{ana3}-2{\rm l(s)}}]}&=&
\frac{8N_{\rm L}^2N_{\rm R}^2}{\left(N_{\rm L}+N_{\rm R}\right)^4}\frac{\Gamma(1-\Gamma)^4}{p_3}
\left(1-{\rm e}^{-p_3\tE/\tD}\right),\nonumber\\
\end{eqnarray} 
where, because of the alignment and proximity of the two encounter stretches, 
the periodic orbit and the other encounter stretch must be backreflected at
the tunnel barriers.  Allowing the ends of the second encounter stretch to
progressively tunnel through into the leads as well, we have, if it tunnels into the lead at one end 
\begin{eqnarray}
\label{li1.17} 
{\rm var}\,G^{[{\ref{ana3}-3{\rm l}}]}&=&
\frac{4\left(N_{\rm L}N_{\rm R}^2+N_{\rm L}^2N_{\rm R}\right)}{\left(N_{\rm L}+N_{\rm R}\right)^3}
\frac{\Gamma^2(1-\Gamma)^3}{p_3}\nonumber\\
&&\times\left(1-{\rm e}^{-p_3\tE/\tD}\right)
\end{eqnarray}
and at both ends
\begin{eqnarray}
\label{li1.16}
{\rm var}\,G^{[{\ref{ana3}-4{\rm l}}]}=
\frac{2N_{\rm L}N_{\rm R}}{\left(N_{\rm L}+N_{\rm R}\right)^2}\frac{\Gamma^3(1-\Gamma)^2}{p_3}
\left(1-{\rm e}^{-p_3\tE/\tD}\right).\nonumber\\
\end{eqnarray}

Such configurations can be also considered for the diagrams in Fig.\
\ref{ana4} and their contributions can be calculated analogously.
To simplify the results, however, we will later perform an expansion
of the contributions in powers of the Ehrenfest time and only retain terms
up to linear order.  As the results for the diagrams of Fig.\ \ref{ana4}
are all of higher order in $\tE$ than the linear one, we will not focus on their explicit
form here.

\subsection{Encounter fringes}

\begin{figure}
\centerline{\includegraphics[width=\columnwidth]{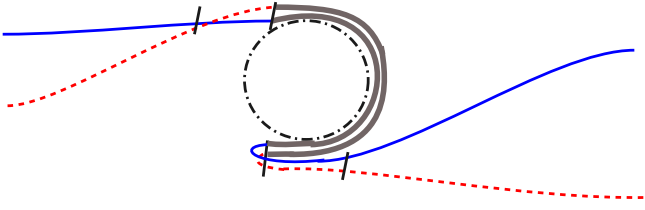}}
\caption{
A periodic orbit encounter with fringe correlations, i.e.\
correlations between the two trajectories that encounter the periodic orbit 
(shown dashed dotted) with each other, but not with the periodic orbit itself. 
The fringe regions are indicated by the black lines perpendicular to the trajectories.
The encounter stretches are shown thick (gray), while the links connecting the encounter
stretches to the opening are indicated by solid (blue) and dashed (red) lines.}
\label{ana5}
\end{figure}

A further effect, and one that actually causes the independence of the
conductance variance of the Ehrenfest time for $\Gamma=1$, are 
correlations during fringes near periodic orbits.
Encounter fringes refer to regions where the two base trajectories which 
encounter the periodic orbit are correlated with each other, but are no 
longer correlated with the periodic orbit itself; see Fig.\ \ref{ana5} for a
schematic depiction.

As the two base trajectories leave the periodic orbit correlated with each other,
we can consider the encounters with the periodic orbit to be aligned.  
The encounter of both trajectories with the periodic orbit has length
$ t_{\mathrm{enc}}$ while the fringes, during which the base trajectories are
correlated have lengths $t_s$ and $t_u$ where the subscripts refer to the 
fact that the stable and unstable distances between the base trajectories must
be small for them to remain correlated. Here, we treat the case in which the 
encounter length is shorter than the periodic orbit. By this we generalize the
calculation of \cite{Bro} to $\Gamma \neq1$.  There, it was shown that the
contribution derived from \eref{li1.2} can be expressed as
\begin{eqnarray}
\label{li1.19} 
{\rm var}\,G^{[{\ref{ana5}}]}&=&
\frac{2N_{\rm L}^2N_{\rm R}^2c^2\lambda}{\left(\pi\hbar\right)^2
\left(N_{\rm L}+N_{\rm R}\right)^4}\int_{1-b}^{b-1}{\rm d}s'{\rm d}u'
\int_0^1{\rm d}u\,u\nonumber\\
&&\times\int_{t_{\rm enc}}^\infty {\rm d}\tau_p
{\rm e}^{-\left[\left(p_3-\Gamma\right){t_{\rm enc}}+\Gamma\tau_p\right]/\tD}
\nonumber\\&&
\times\left(\left|\frac{s'}{b-1}\right|^{\frac{p_2}{\lambda\tD}}-1\right)
\left(\left|\frac{u'}{b-1}\right|^{\frac{p_2}{\lambda\tD}}-1\right)\nonumber\\
&&\times\cos\left[\frac{c^2u\left(s'-u'\right)}{\hbar}\right]
\end{eqnarray}
where $b$ is again a classical constant of order unity. The stable and unstable coordinates are defined as in \cite{Bro}: As correlation effects away from the periodic orbits get only important when the encountering orbits approach and leave this at classically close points one first characterize the correlation of all three orbit pieces (i.e.\ the periodic orbit and the two encountering orbits) by the coordinate $u$. Furthermore the difference between the stable and the unstable  coordinates of the encountering orbits at the beginning and the end of the encounter is denoted by $s'$ and $u'$. These coordinates characterize the time the encountering orbits remain correlated before and after they are correlated with the periodic orbit, respectively.  
In \eref{li1.19}, we included the correct dwell times for the different parts of the
trajectory: $\tD/\Gamma$ for the isolated links, $\tD/p_2$ for the fringes and
$\tD/p_3$ for the encounter stretches correlated with the periodic orbit.
Following the steps in \cite{Bro}, this contribution evaluates to
\begin{eqnarray}
\label{li1.20} 
{\rm var}\,G^{[{\ref{ana5}}]}&=&
\frac{2N_{\rm L}^2N_{\rm R}^2}{\left(N_{\rm L}+N_{\rm R}\right)^4}
\frac{p_2^2}{\Gamma\left(2p_2-p_3\right)}\nonumber\\
&&\times\left({\rm e}^{-p_3\tE/\tD}-{\rm e}^{-2p_2\tE/\tD}\right).
\end{eqnarray}
Without tunnel barriers, as shown in \cite{Bro}, further contributions can be obtained 
when the fringes start in the leads.  Similarly, with tunnel barriers, we can
have the fringes tunnel into the leads.  This leads to an additional time integral
over the duration of the fringe that remains inside the system.  When one fringe tunnels
into the lead, two links are removed and the contribution becomes
\begin{eqnarray}
\label{li1.21} 
{\rm var}\,G^{[{\ref{ana5}-2{\rm l(s)}}]}&=&
-\frac{2\left(N_{\rm L}^2N_{\rm R}+N_{\rm L}N_{\rm R}^2\right)}{\left(N_{\rm L}+N_{\rm R}\right)^3}
\frac{\Gamma p_2}{\left(2p_2-p_3\right)}\nonumber\\
&&\times\left({\rm e}^{-p_3\tE/\tD}-{\rm e}^{-2p_2\tE/\tD}\right).
\end{eqnarray}
The `(s)' in the superscript refers to the fact that the links were removed from
the same side of the encounter. When both fringes, and all four links are removed,
we have
\begin{eqnarray}
\label{li1.22} 
{\rm var}\,G^{[{\ref{ana5}-4{\rm l}}]}&=&
\frac{2N_{\rm L}N_{\rm R}}{\left(N_{\rm L}+N_{\rm R}\right)^2}
\frac{\Gamma^3}{\left(2p_2-p_3\right)}\nonumber\\
&&\times\left({\rm e}^{-p_3\tE/\tD}-{\rm e}^{-2p_2\tE/\tD}\right).
\end{eqnarray}

With the tunnel barriers though, new possibilities also arise.  For example, when
a fringe tunnels into the lead, one of the trajectories could be backscattered,
as for the 2-encounters previously.  Similarly to \eref{li0.4}, we obtain here
\begin{eqnarray}
\label{li1.225} 
{\rm var}\,G^{[{\ref{ana5}-1{\rm l}}]}&=&
-\frac{8N_{\rm L}^2N_{\rm R}^2}{\left(N_{\rm L}+N_{\rm R}\right)^4}
\frac{(1-\Gamma)p_2}{\left(2p_2-p_3\right)}\nonumber\\
&&\times\left({\rm e}^{-p_3\tE/\tD}-{\rm e}^{-2p_2\tE/\tD}\right).
\end{eqnarray}
Furthermore we can move additionally one fringe into the lead yielding
\begin{eqnarray}
\label{li1.2255} 
{\rm var}\,G^{[{\ref{ana5}-3{\rm l}}]}&=&
\frac{4\left(N_{\rm L}N_{\rm R}^2+N_{\rm L}^2N_{\rm R}\right)}{\left(N_{\rm L}+N_{\rm R}\right)^3}
\frac{(1-\Gamma)\Gamma^2}{\left(2p_2-p_3\right)}\nonumber\\
&&\times\left({\rm e}^{-p_3\tE/\tD}-{\rm e}^{-2p_2\tE/\tD}\right).
\end{eqnarray}
If both fringes are partially backscattered at the leads, we have the contribution
\begin{eqnarray}
\label{li1.226} 
{\rm var}\,G^{[{\ref{ana5}-2{\rm l}}]}&=&
\frac{8N_{\rm L}^2N_{\rm R}^2}{\left(N_{\rm L}+N_{\rm R}\right)^4}\frac{\Gamma(1-\Gamma)^2}
{\left(2p_2-p_3\right)}\nonumber\\&&
\times\left({\rm e}^{-p_3\tE/\tD}-{\rm e}^{-2p_2\tE/\tD}\right).
\end{eqnarray}

Next we can allow the periodic orbit itself, rather than the fringes, to touch the lead and be reflected from the tunnel barriers.  The encounter stretches may then both tunnel into the lead, or one may be reflected.  We recall that the case in which  both are reflected is already included in the survival probability of the encounter. 
One fringe is therefore removed from the systems, while the fringe that remains could then be inside the system, tunnel into the lead or be partially backreflected  at the opening.  If the periodic orbit touches both leads during the encounter, we return to the situation in the previous subsection, so here we only consider the cases in which at least some of the second fringe remains.  

The removal of one fringe entirely implies the following changes in \eref{li1.19}:  With only one fringe time left, we only have one of the factors in the third line of \eref{li1.19}.  Also the encounter time is replaced by an integration variable (with a range from $0$ to $t_{\rm enc}$) which accounts for the amount of the encounter left after the periodic orbit is reflected from the tunnel barriers. Performing, with these changes, again the steps in \cite{Bro} analogous to those  following \eref{li1.19} above, we obtain the contributions for the various different cases.  First, when the remaining fringe lies inside the system, we could have the encounter (which touches the lead) partially reflected
\begin{eqnarray}
\label{li1.227} 
{\rm var}\,G^{[{\ref{ana5}-1{\rm f}-1{\rm l}}]}&=&
-\frac{4N_{\rm L}^2N_{\rm R}^2}{\left(N_{\rm L}+N_{\rm R}\right)^4}\frac{(1-\Gamma)^2p_2}{p_3}\nonumber\\
&&\times\left(1-{\rm e}^{-p_3\tE/\tD}\right){\rm e}^{-p_2\tE/\tD},
\end{eqnarray}
where `-1f-1l' refers to the fact that one fringe and one link were removed. Likewise, both encounter stretches could tunnel directly into the lead so that two links are removed on the same side of the encounter
\begin{eqnarray}
\label{li1.228} 
{\rm var}\,G^{[{\ref{ana5}-1{\rm f}-2{\rm l(s)}}]}&=&
-\frac{\left(N_{\rm L}N_{\rm R}^2+N_{\rm L}^2N_{\rm R}\right)}{\left(N_{\rm L}+N_{\rm R}\right)^3}
\frac{(1-\Gamma)p_2\Gamma}{p_3}\nonumber\\
&&\times\left(1-{\rm e}^{-p_3\tE/\tD}\right){\rm e}^{-p_2\tE/\tD}.
\end{eqnarray}
The remaining fringe could also be paritally reflected from the tunnel barriers. For the encounter stretches have the same two possibilities as above.  This leads to
\begin{eqnarray}
\label{li1.229} 
{\rm var}\,G^{[{\ref{ana5}-1{\rm f}-2{\rm l}}]}&=&
\frac{8N_{\rm L}^2N_{\rm R}^2}{\left(N_{\rm L}+N_{\rm R}\right)^4}\frac{(1-\Gamma)^3\Gamma}{p_3}\nonumber\\
&&\times\left(1-{\rm e}^{-p_3\tE/\tD}\right){\rm e}^{-p_2\tE/\tD} ,
\end{eqnarray}
and
\begin{eqnarray}
\label{li1.2295} 
{\rm var}\,G^{[{\ref{ana5}-1{\rm f}-3{\rm l(s)}}]}&=&
\frac{2\left(N_{\rm L}^2N_{\rm R}+N_{\rm L}N_{\rm R}^2\right)}{\left(N_{\rm L}+N_{\rm R}\right)^3}
\frac{(1-\Gamma)^2\Gamma^2}{p_3}\nonumber\\
&&\times\left(1-{\rm e}^{-p_3\tE/\tD}\right){\rm e}^{-p_2\tE/\tD}.
\end{eqnarray}
Finally the remaining fringe can tunnel directly into the lead, giving the contributions
\begin{eqnarray}
\label{li1.2296} 
{\rm var}\,G^{[{\ref{ana5}-1{\rm f}-3{\rm l}}]}&=&
\frac{2\left(N_{\rm L}^2N_{\rm R}+N_{\rm L}N_{\rm R}^2\right)}{\left(N_{\rm L}+N_{\rm R}\right)^3}
\frac{(1-\Gamma)^2\Gamma^2}{p_3}\nonumber\\
&&\times\left(1-{\rm e}^{-p_3\tE/\tD}\right){\rm e}^{-p_2\tE/\tD} ,
\end{eqnarray}
and
\begin{eqnarray}
\label{li1.2297} 
{\rm var}\,G^{[{\ref{ana5}-1{\rm f}-4{\rm l}}]}&=&
\frac{2N_{\rm L}N_{\rm R}}{\left(N_{\rm L}+N_{\rm R}\right)^2}
\frac{(1-\Gamma)^2\Gamma^3}{p_3}\nonumber\\
&&\times\left(1-{\rm e}^{-p_3\tE/\tD}\right){\rm e}^{-p_2\tE/\tD}.
\end{eqnarray}

\subsection{Multiple periodic orbit traversals}

Having treated all encounter configurations with periodic orbits where the
encounter stretches are shorter than the enclosed periodic orbit we now turn to
the corresponding contributions where the encounter stretches are longer than
the enclosed periodic orbit. In this context only diagrams with stretches that
both possess the same number of traversals around the enclosed periodic orbit
yield a contribution, for a justification see \cite{Bro}. In this context we
consider that each encounter stretch has $k$ full windings around the enclosed
periodic orbit. We take here the two encounter times $t_{{\rm enc},1}$,
$t_{{\rm enc},2}$ and the corresponding primitive times $t_{{\rm enc},i}^p\equiv
t_{{\rm enc},i}-k\tau_p$. These primitive times are again shorter than the
enclosed periodic orbit and we can now also consider the different cases
depicted in Figs.\ \ref{ana2}--\ref{ana5}, and how they can be arranged. 
We start with the case in which the two primitive encounter times, $t_{{\rm enc},i}^p$, 
do not overlap.  When $k=0$, the corresponding diagram is depicted in Fig.\ \ref{ana2}.
To explain the calculation we begin with \Eref{li1.2} for $k=0$ and perform the
integrals over the links $t_{i}$, and over $t'$ where the latter integral leads to the factor 
$\left(\tau_p-t_{{\rm encs}}\right)$:
\begin{eqnarray}
\label{li1.23}
{\rm var}\,G^{[{\ref{ana2}}]}_{k=0}&=&
\frac{N_{\rm L}^2N_{\rm R}^2}{\left(N_{\rm L}+N_{\rm R}\right)^{4}}
\int_{-c}^c {\rm d}\boldsymbol{s}{\rm d}\boldsymbol{u}
\frac{\TH^2}{\Omega^2}\frac{{\rm e}^{\frac{{\rm i}}{\hbar}\boldsymbol{s}\boldsymbol{u}}}
{t_{{\rm enc},1} t_{{\rm enc},2}} \nonumber \\
&&\times\int_{t_{{\rm encs}}}^\infty 
{\rm d}\tau_p\left(\tau_p-t_{{\rm encs}}\right) \nonumber \\
&& \times {\rm e}^{-\left[\left(p_2-\Gamma\right)t_{{\rm encs}}
+\Gamma \tau_p\right]/\tD} .
\end{eqnarray}

To turn to $k\neq0$, $t_{{\rm enc},i}$ is then replaced only inside the integrand of the
$\tau_p$-integral by $t_{{\rm enc},i}^p$, these times are shorter than 
$\tau_p$ by definition. The $t_{{\rm enc},i}$ before the $\tau_p$-integral compensate the overcounting of equivalent positions of the Poincar\'e surface of sections by the $s_i, u_i$-integrals and thus do not need to be altered. The possibilities for placing these stretches around the 
periodic orbit are the same as they were for $t_{{\rm enc},i}$ for $k=0$. 
This allows us to treat $k>0$ in essentially the same way we treated $k=0$ before.
As $t_{{\rm enc},i}^p= t_{{\rm enc},i}-k\tau_p$, we reexpress the primitive
encounter times in terms of $t_{{\rm enc},i}$ and $\tau_p$.
The limits of the $\tau_p$-integration are also altered for $k\neq0$: As the
primitive encounters do not (yet) overlap, we have the condition 
$t_{{\rm enc},1}^p+t_{{\rm enc},2}^p\geq\tau_p$ so that the
lower limit is now $t_{{\rm encs}}/(2k+1)$.
Since we remove all $k$ complete windings of the periodic orbit from both 
encounter times, the shorter encounter must be $t_{\rm enc, min}\geq k\tau_p$
so that the upper limit is $t_{\rm enc, min}/k$.  Finally $p_j$ is replaced by
$p_{2k+j}$ (where in this context $\Gamma$ is defined as $p_1$) since during 
the primitive encounter stretches we have a total of $2k+2$ correlated stretches
while elsewhere we have $2k+1$.  All these replacements finally yield
\begin{eqnarray}
\label{li1.23.5}
{\rm var}\,G^{[{\ref{ana2}}]}_{{\rm mt}}&=&
\frac{N_{\rm L}^2N_{\rm R}^2}{\left(N_{\rm L}+N_{\rm R}\right)^{4}}
\int_{-c}^c {\rm d}\boldsymbol{s}{\rm d}\boldsymbol{u}
\frac{\TH^2}{\Omega^2}\frac{{\rm e}^{\frac{{\rm i}}{\hbar}\boldsymbol{s}\boldsymbol{u}}}
{t_{{\rm enc},1} t_{{\rm enc},2}} \\
&&\times\int_{\frac{t_{{\rm encs}}}{2k+1}}^{\frac{t_{{\rm enc,min}}}{k}} 
{\rm d}\tau_p\left[(2k+1)\tau_p-t_{{\rm encs}}\right]\nonumber\\
&&\times{\rm e}^{-\left[\left(p_{2k+2}-p_{2k+1}\right)
\left(t_{{\rm encs}}-2k\tau_p\right)+p_{2k+1}\tau_p\right]/\tD} , \nonumber
\end{eqnarray}
where the additional index `mt' indicates that this contribution results from
multiple traversals $k>0$. Performing now the remaining integrals in the same way 
as for example in Eqs.\ (\ref{li1.4}--\ref{li1.7}) we obtain this contribution.
As the full expression is rather involved we just give here terms up to 
linear order in $\tE$:
\begin{eqnarray}
\label{li1.24}
{\rm var}\,G^{[{\ref{ana2}}]}_{\rm mt} &=&
-\frac{2N_{\rm L}^2N_{\rm R}^2}{\left(N_{\rm L}+N_{\rm R}\right)^4}\sum_{k=1}^\infty
\frac{\left(p_{2k+2}-p_{2k+1}\right)\tE}{k^2(2k+1)\tD} \nonumber \\
&&{}+{\mathcal O}\left(\tE^2\right) .
\end{eqnarray}
We also omitted here terms that are also nonzero for $p_k=1$ as they all cancel with the contributions below.
Also in this case one link can be removed leading to a configuration where one
encounter stretch touches the opening. This implies that in \eref{li1.23.5}
one factor $\tau_D/\Gamma$ is removed, the duration of the encounter
stretch which touches the tunnel barrier is replaced by an integration variable
integrated from $0$ to $t_{{\rm enc},i}$, and additionally a factor 
$(1-\Gamma)^{2k+1}$ arises to account for the probability that all the $2k+1$ orbital 
parts surrounding the periodic orbit are backreflected when they hit the lead. 
This contribution then yields
\begin{eqnarray}
\label{li1.25}
{\rm var}\,G^{[{\ref{ana2}-1l}]}_{\rm mt}&=&
\frac{4N_{\rm L}^2N_{\rm R}^2}{\left(N_{\rm L}+N_{\rm R}\right)^4}\sum_{k=1}^\infty
\frac{\left(p_{2k+2}-p_{2k+1}\right)\tE}{k^2(2k+1)\tD}\nonumber\\
&&{}+{\mathcal O}\left(\tE^2\right). 
\end{eqnarray}
Shrinking two links connecting two \textit{different} encounter stretches to the
opening, we obtain (following the same steps as just described also for the other $t_{{\rm enc},i}$) no contribution 
which is linear in $\tau_E$.   Shrinking two links connecting the \textit{same} 
encounter we need again to introduce in \eref{li1.23.5} two additional time integrals 
as described before \eref{li1.13}. This also finally yields zero contribution linear in $\tE$, because 
the terms resulting from the two limits of the $\tau_p$-integral cancel. The same also holds for contributions following.

Similar contributions are also obtained in the other cases. For a generalized
3-encounter introduced around \Eref{li1.75} we obtain the integrals given
in \eref{ap2} which provide the contribution
\begin{eqnarray}
\label{li1.26}
{\rm var}\,G^{[{\ref{ana3}}]}_{\rm mt}&=&
\frac{2N_{\rm L}^2N_{\rm R}^2}{\left(N_{\rm L}+N_{\rm R}\right)^4}\sum_{k=1}^\infty
\frac{\tE}{k(k+1)\tD} \\
&&\times \left(2p_{2k+2}-p_{2k+1}-p_{2k+3}\right)
+{\mathcal O}\left(\tE^2\right) , \nonumber
\end{eqnarray}
when the encounters lie inside the system and
\begin{eqnarray}
\label{li1.27}
{\rm var}\,G^{[{\ref{ana3}-1{\rm l}}]}_{\rm mt}&=&
-\frac{4N_{\rm L}^2N_{\rm R}^2}{\left(N_{\rm L}+N_{\rm R}\right)^4}\sum_{k=1}^\infty
\frac{\tE}{k(k+1)\tD} \\
&&\times\left(2p_{2k+2}-p_{2k+1}-p_{2k+3}\right)
+{\mathcal O}\left(\tE^2\right) , \nonumber 
\end{eqnarray}
when one link is removed. Additionally we can shrink both links connecting the
3-encounter to the opening on one side.  This amounts to removing two links, and
introducing an additional integral over the part of the encounter that remains
inside the system, in \eref{ap2}, yielding
\begin{eqnarray}
\label{li1.28}
{\rm var}\,G^{[{\ref{ana3}-2{\rm l}}]}_{\rm mt}&=&
\frac{2\left(N_{\rm L}N_{\rm R}^2+N_{\rm L}^2N_{\rm R}\right)}{\left(N_{\rm L}+N_{\rm R}\right)^3}
\sum_{k=1}^\infty\frac{\tE}{k(k+1)\tD} \\
&&\times\left(2p_{2k+2}-p_{2k+1}-p_{2k+3}\right)
+{\mathcal O}\left(\tau_E^2\right) . \nonumber
\end{eqnarray}
Again zero contribution is obtained when more links are removed.

For a non-aligned 3-encounter, when the encounter lies inside the
system we obtain from Eqs.\ (\ref{ap31},\ref{ap32})
\begin{eqnarray}
\label{li1.29}
{\rm var}\,G^{[{\ref{ana4}({\rm a})}]}_{\rm mt}&=&
\frac{4N_{\rm L}^2N_{\rm R}^2}{\left(N_{\rm L}+N_{\rm R}\right)^4}\sum_{k=1}^\infty
\left[\frac{\left(p_{2k+3}-p_{2k+2}\right)\tE}{(k+1)(2k+1)\tD}-\right.\nonumber\\
&&\left.-\frac{\left(p_{2k+2}-p_{2k+1}\right)\tE}{k(2k+1)\tD}\right]
+{\mathcal O}\left(\tE^2\right) 
\end{eqnarray}
and when one link is removed:
\begin{eqnarray}
\label{li1.30}
{\rm var}\,G^{[{\ref{ana4}({\rm a})-1{\rm l}}]}_{\rm mt}&=&
-\frac{8N_{\rm L}^2N_{\rm R}^2}{\left(N_{\rm L}+N_{\rm R}\right)^4}\sum_{k=1}^\infty
\left[\frac{\left(p_{2k+3}-p_{2k+2}\right)\tE}{(k+1)(2k+1)\tD}\right.\nonumber\\
&&+\frac{\left(2p_{2k+2}-p_{2k+1}-p_{2k+3}\right)\tE}{2k(k+1)\tD}\nonumber\\
&&\left.-\frac{\left(p_{2k+2}-p_{2k+1}\right)\tE}{k(2k+1)\tD}\right]
+{\mathcal O}\left(\tE^2\right).
\end{eqnarray}
Finally we treat the configuration in which the encounter stretches overlap at both
ends.  When $k=0$ this configuration is depicted in Fig.\ \ref{ana4}(b). 
With multiple traversals of the periodic orbit, the contribution given in \Eref{ap4}
provides, when the stretches lie inside the system,
\begin{eqnarray}
\label{li1.31}
{\rm var}\,G^{[{\ref{ana4}({\rm b})}]}_{\rm mt}&=&
-\frac{2N_{\rm L}^2N_{\rm R}^2}{\left(N_{\rm L}+N_{\rm R}\right)^4}\sum_{k=1}^\infty
\frac{\left(p_{2k+3}-p_{2k+2}\right)\tE}{(k+1)^2(2k+1)\tD}\nonumber\\
&&+{\mathcal O}\left(\tE^2\right)
\end{eqnarray}
and when one link is removed
\begin{eqnarray}
\label{li1.32}
{\rm var}\,G^{[{\ref{ana4}({\rm b})-1{\rm l}}]}_{\rm mt}&=&
\frac{4N_{\rm L}^2N_{\rm R}^2}{\left(N_{\rm L}+N_{\rm R}\right)^4}\sum_{k=1}^\infty
\frac{\left(p_{2k+3}-p_{2k+2}\right)\tE}{(k+1)^2(2k+1)\tD}\nonumber\\
&&+{\mathcal O}\left(\tE^2\right).
\end{eqnarray}
No further contributions are obtained when taking into account fringes as in
Fig.\ \ref{ana5}.  This can be checked from \Eref{li1.19} by adjusting the limits of the
$\tau_p$-integral appropriately.  Then, the terms from the two limits of the
$\tau_p$-integral cancel to linear order in $\tE$.

\subsection{Linear Ehrenfest-time dependence}

Having calculated all the contributions we can now sum them to obtain the
overall contributions to the conductance variance, to leading order in the 
number of open channels in the leads.  First we can check that all the terms after 
Sec.\ \ref{discrete} are zero for $\tE=0$, so that the RMT result \eref{li22.1} is preserved.
The case that is especially interesting for the comparison with the
numerics is the contribution to the conductance variance
when $N_{\rm L}=N_{\rm R}$, for arbitrary $\Gamma$ and of linear order in $\tE$.
We therefore consider for ${N_{\rm L}=N_{\rm R}}\rightarrow \infty$, in the presence of time reversal symmetry,
\begin{eqnarray}
\label{li22.9}
{\rm var}\,G_{N_{\rm L}=N_{\rm R}}&\approx&{\rm var}\,G^{\rm RMT}+\alpha\frac{\tE}{\tD}\nonumber\\&\approx& \frac{1}{8}\left[1+\left(1-\Gamma\right)^2\right]+{\alpha}\frac{\tE}{\tD}.
\end{eqnarray}
We obtain 
\begin{eqnarray}
\label{li23}
{\alpha}(\Gamma)&=&\frac{\Gamma}{4}\left[\left(\Gamma-1\right)\left(7\Gamma^3-6\Gamma^2+4\Gamma-2\right)
+\ln \left(2\Gamma-\Gamma^2\right)\right.\nonumber\\
&&\left.\times\frac{\Gamma^2\left(2-\Gamma\right)}{(1-\Gamma)}
-{\Gamma}{\rm Li}_2\left((1-\Gamma)^2\right)\right] 
\end{eqnarray}
by summing the contributions linear in $\tau_E$ from Eqs.\ (\ref{li0.3}-\ref{li0.91}, \ref{li1}, \ref{li1.76}-\ref{li1.8}, \ref{li1.10}-\ref{li1.12}, \ref{li1.14}-\ref{li1.16}, \ref{li1.20}-\ref{li1.2297}, \ref{li1.24}-\ref{li1.32}).  
Here ${\rm Li}_2(x)$ denotes the polylogarithmic function, which arises when performing the
$k$-summations above. The function ${\alpha}(\Gamma)$ is shown as the solid (red) line in Fig.\ \ref{condfluc}. 


Having derived the complete set of contributions, we now discuss the main
terms that lead to the two extrema in Fig.\ \ref{condfluc}. 
First we note that the contributions from multiply-traversed periodic orbits
are quite small. When $\Gamma=1$ they provide no contribution linear in
$\tE$, as can be checked from Eqs.\ (\ref{li1.24}-\ref{li1.32}). For most of the contributions for $\Gamma\neq1$ the index $k$ of the $p_k$'s in the sum is much larger than one. In this case the differences of the $p_k$'s in Eqs.\ (\ref{li1.24}-\ref{li1.32}) tend to one and thus the contributions from these equations tend to zero. Then the result is almost equal to the vanishing value for $\Gamma=1$.

In Fig.\ \ref{condfluc}, the maximum lies in a region of small $\Gamma$. In this regime, most of the diagrams approximately
cancel each other as can be seen by expanding their contributions in a Taylor series for
small $\Gamma$: The ones from diagrams containing encounters inside the
system [e.g.\ the contribution \eref{li0.3}] are canceled by the contributions obtained
when one and two links are shrunk [e.g.\ the contributions
(\ref{li0.4}--\ref{li0.7})].  However, this cancellation does not hold for the
3-encounter with the periodic orbit (see Fig.\ \ref{ana3}). In this case we have
two possibilities: First that the encounter lies inside the system or touches
the lead at one end [see Eqs.\ (\ref{li0.8}--\ref{li0.91})]. 
In total this leads to an Ehrenfest-time dependent factor of ${\rm e}^{-p_3\tau_E/\tau_D}$.
Additionally, we also need to take into account the contributions from
encounters touching the leads on both sides leading to an increasing
contribution proportional to $1-{\rm e}^{-p_3\tau_E/\tau_D}$ with a larger
prefactor than the first one [see Eqs.\ (\ref{li1.18}--\ref{li1.16})]. 
Together these cause the first peak in Fig.\ \ref{condfluc}. 

In the case of the dip the factor $(1-\Gamma)$, 
i.e.\ the probability for the particle to be backreflected
at the opening, is quite small. This implies that we get the main contributions
from diagrams with encounters inside the system (i.e.\ the ones also obtained
for $\Gamma=1$) depicted in Figs.\ \ref{ana1}--\ref{ana5}.  These together yield
a negative contribution. 

\section{Numerical simulation} \label{part2}

We numerically confirm our main analytical prediction, Eqs.~(\ref{li22.9})
and (\ref{li23}). The model we use is the open kicked rotator
with time-dependent Hamiltonian~\cite{Izr}
\begin{equation}
\label{num0}
\hat H=\frac{\left(p+p_0\right)^2}{2}
+K\cos\left(x+x_0\right)\sum_{n=-\infty}^\infty
\delta\left(t-n\tau_f\right) ,
\end{equation}
with $\tau_f$ the free flight time. Depending on the kicking strength $K$ the
dynamics changes from integrable for $K=0$ to fully chaotic for $K\gtrsim 7$. In the
latter regime, local exponential instability is characterized by the
Lyapunov exponent
\begin{equation}
\label{num1}
\lambda=\frac{1}{\tau_f}\ln\left(\frac{K}{2}\right).
\end{equation}

\begin{figure}
\centerline{\includegraphics[width=0.47\textwidth]{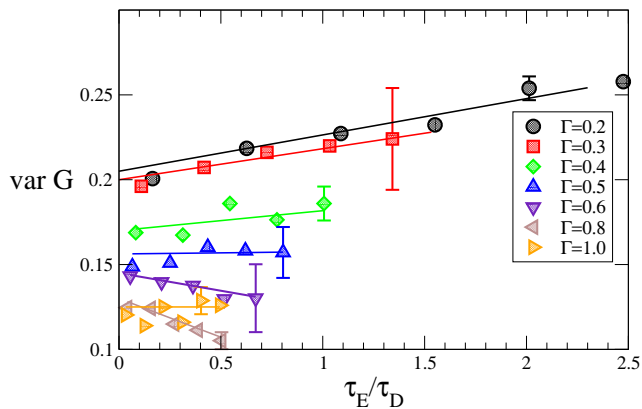}}
\caption{Conductance variance ${\rm var}\,G(E)$ as a function of $\tE/\Gamma \tD$ for
various values of $\Gamma$. The straight lines are linear fits constrained to go to 
$(1+(1-\Gamma)^2)/8$ at $\tE=0$. Their slope give the parameter $\alpha$ defined in 
Eqs.\ (\ref{li22.9}) and (\ref{li23}).}
\label{condflucraw}
\end{figure}

The quantities $p_0$ and $x_0$ are introduced to break the Hamiltonian's two
symmetries \cite{Izr}, and investigate different symmetry classes with ($p_0$ or $x_0=0$) and
without ($p_0 \ne 0 \ne x_0$) time reversal symmetry. 
The Ehrenfest time in this system is determined, within a constant of little relevance, by
$\tE\equiv \lambda^{-1} \ln(M/c)$ with $M$ the size of the Hilbert space, determined by the
quantization of the Hamiltonian via discretization of the 
coordinates as, e.g.  $p_l=2 \pi l/M$, $l=1,\ldots M$. Here $c$ is a system-dependent
constant of order one
that is of classical origin, and as such does not depend on $M$.
A quantum representation of the Hamiltonian (\ref{num0}) is
then provided by the unitary $M \times M$ 
Floquet matrix $U$, 
giving  the time evolution for one iteration of the map defined by $\hat H$ in a time
interval $[t_0,t_0+\tau_f[$. 
For our specific choice of the kicked rotator,
the Floquet operator has matrix elements
\begin{eqnarray}\label{kickedU}
U_{l,l'} &=& M e^{-(\pi i/M) [(l+l_0)^{2}+(l'+l_0)^2]}
\\
& \times & \sum_m e^{2 \pi i m(l-l')/M}   
e^{-(iMK/4\pi) \cos(2\pi (m+m_0)/M)} \nonumber
\end{eqnarray}
with $l_0=p_0 M/2 \pi$ and $m_0=x_0 M/2 \pi$.
Transport can finally be investigated once absorbing phase-space strips are introduced
to model contacts to leads. This is achieved by means of projection operators $P$.
In our case of tunnel-coupled leads, the latter are $N \times M$ diagonal matrices with
entries $P_{ij}=\delta_{ij} \sqrt{\Gamma}$, assuming that the system is coupled to 
all $N=N_{\rm L}+N_{\rm R}$ channels with the same transparency $0<\Gamma \le 1$.
The scattering matrix is finally defined as~\cite{Fyodorov}
\begin{equation}
S(\varepsilon) = (1-P^T P)^{1/2} - P [\exp(-i \varepsilon) - U (1-P^T P)]^{-1} U P^T \, .
\end{equation}
The kicked rotator model is particularly performant to investigate the semiclassical limit
with $\tE \gtrsim \tD$, as it allows for a rather large variation of the system
size and hence the Ehrenfest time. Previous such investigations of the open
kicked rotator are presented e.g. in Refs.~\cite{Tworzyd,Suk,Schom,Jacquod}, and we refer
the reader to these works for further details of the model. The variance of the conductance
is calculated by varying the quasienergies 
$\varepsilon$ at given lead position. Averages of ${\rm var}\,G(E)$ are further performed 
with different lead positions.

We first show in Fig.~\ref{condflucraw} the behavior of ${\rm var}\, G(E)$ as a function of 
$\tE/\tau_D$. We clearly see $\Gamma$-dependent behaviors, as ${\rm var}\,G(E)$ increases
for small $\Gamma$ and decreases for larger $\Gamma$ until it becomes
independent of $\tE$ for $\Gamma=1$. We extract the linear slope of the
${\rm var}\, G(E)$ vs. $\tE/\tau_D$ curve via a linear fit, setting $\tE=0$ where the curves
intersect the RMT universal value ${\rm var} (G)^{\rm RMT} = (1+(1-\Gamma)^2)/8$~\cite{BroRMT}.
Further dividing this slope by $\Gamma$ gives us $\alpha$. 
There is still an uncertainty in $\alpha$ due to an uncertainty in the 
precise value of the Lyapunov exponent -- the latter has been found numerically to 
deviate in open systems from its exact value (\ref{num1}), which has been attributed to
finite-time effects -- and because of the arbitrary constant $c$ of order one in the definition 
of $\tE$. We remove this uncertainty by forcing 
numerical and analytical data to agree for $\Gamma=0.2$. Once this is done, there is no
free parameter left. We compare the so obtained
numerical values for $\alpha$ with the analytical prediction of Eq.~(\ref{li23}) in Fig.~\ref{condfluc}.
The excellent agreement between numerical data and the analytical curve fully
confirms our theory. We also checked, but do not show, that the linear $\tE$-dependence of
${\rm var }\, G(E)$ is halved when time-reversal symmetry is broken.

\section{Conclusions} \label{concs}

In this article we determined the dependence of the conductance variance
of a chaotic cavity with tunnel barriers on the tunneling probability $\Gamma$.  In particular we find an Ehrenfest-time dependence for the general case $\Gamma<1$ of non-perfect coupling. 
We focused on the contribution which is linear in the Ehrenfest time and at leading
order in the total number of open channels $N$.  We predict a
nonmonotonous sinusoidal behavior with one maximum at $\Gamma\approx0.2$ and 
a (roughly 50\%) deeper minimum at $\Gamma\approx0.8$.  This analytical result was derived 
semiclassically by systematically considering a rather large number of possible configurations, 
but the general behavior derives from a much smaller set.  We discussed the 
main contributions which lead to each of the two peaks to provide a better intuitive 
understanding of the overall structure. Finally, we compared these analytical 
predictions with numerical simulations performed for the kicked rotor (in the 
chaotic regime).  There was good agreement between the two, with the analytical curve
within the error bars of all the numerically determined datapoints. 

Although, even at leading order, we treated a large number of possible
semiclassical diagrams, they could be constructed, as we showed in this article, in
a controlled way from the diagrams that exist when the tunnel barriers are absent.
This calculation could therefore possibly be extended to higher
order contributions in $1/N$ and $\tE$, but a more natural next step would be
to extend the Ehrenfest-time dependence of the (leading order in $N$) 
results for the full counting statistics of transport moments \cite{Wal} to include
tunnel barriers in the leads and to account for the extra diagrammatic possibilities
that then arise.

On the numerical side, we emphasize that the agreement observed in this article
is the first observed for the conductance properties of systems with tunnel barriers.
Neither the results for the conductance nor for shot noise obtained in
\cite{Whitney} could yet be confirmed by numerical simulations. Further
numerical investigation in this area would therefore be highly desirable. 
Finally it would be interesting to check the observed phenomena experimentally. 
Although Ehrenfest-time effects in antidot superlattices were observed more than ten
years ago \cite{Yevtu}, none of the other Ehrenfest-time dependencies predicted for chaotic
systems could be checked so far experimentally.  This has been mainly due to the small 
range through which the system size can be varied in quantum dots.  However, with the 
additional parameter of the tunneling probability and in view of the expected double peak structure, the 
prediction in this article could be amenable to experimental verification.

\section*{Acknowledgements}

This work was funded by the Deutsche Forschungsgemeinschaft (Research Unit FOR 760) (DW and KR) and by the Alexander von Humboldt foundation (JK). PJ has been partially supported by the NSF 
under grant No DMR-0706319 and by the Swiss Center of Excellence MaNEP.  
It is a great pleasure to thank C.\ Petitjean for discussions and for drawing our attention to this problem. Furthermore we thank R.\ Whitney for discussions at the early stage of this project.

\appendix

\section{Periodic orbit encounters}

Here we give the necessary integrals for obtaining the contributions from diagrams containing enclosed
periodic orbits that are surrounded $k$-times by each encounter stretch. In
these expressions the link and $t'$-integrals in \Eref{li1.2} have already
been performed, but not yet the $\tau_p$- and the $\boldsymbol{s},\boldsymbol{u}$-integrals.
These expressions are especially useful because they not only allow one
to obtain the contributions from encounters inside the system but
(via integrations with respect to the corresponding encounter times) also the
contributions when encounters touch the openings.

The corresponding expression obtained for two independent 2-encounters was already given in the main text in Eq.\ (\ref{li1.23.5}). In the case of a generalized 3-encounter we obtain
\begin{eqnarray}
\label{ap2}
{\rm var}\,G^{[{\ref{ana3}}]}_{\rm mt}&=&
\frac{2N_{\rm L}^2N_{\rm R}^2}{\left(N_{\rm L}+N_{\rm R}\right)^4}
\int_{-c}^c{\rm d}\boldsymbol{s}{\rm d}\boldsymbol{u}
\int_{\frac{t_{{\rm enc,max}}}{k+1}}^{\frac{t_{{\rm
enc,min}}}{k}}{\rm d}\tau_p\frac{\TH^2}{\Omega^2}\nonumber\\
&&\times\frac{{\rm e}^{\frac{{\rm i}}{\hbar}\boldsymbol{s}\boldsymbol{u}}}
{t_{{\rm enc},1}t_{{\rm enc},2}}\left(t_{\rm enc,max}-t_{\rm enc,min}\right)\nonumber\\
&&\times{\rm e}^{-\left(p_{2k+3}-p_{2k+2}\right)\left(t_{\rm enc,min}-k\tau_p\right)/\tD}\\
&&\times{\rm e}^{-\left[\left(p_{2k+2}-p_{2k+1}\right)\left(t_{\rm enc,max}-k\tau_p\right)+p_{2k+1}\tau_p\right]/\tD}\nonumber.
\end{eqnarray}
For a non-aligned 3-encounter, similar to (\ref{li1.4},\ref{li1.5}), it is given
first for $\tau_p>t_{{\rm enc},1}^p+t_{{\rm enc},2}^p$ by
\begin{eqnarray}
\label{ap31}                    
{\rm var}\,G^{[{\ref{ana4}({\rm a})}]}_{\rm mt(1)}&=&
\frac{2N_{\rm L}^2N_{\rm R}^2}{\left(N_{\rm L}+N_{\rm R}\right)^4}\frac{2\tD}{\left(p_{2k+3}+p_{2k+1}-2p_{2k+2}\right)}\nonumber \\
&&\times
\int_{-c}^c{\rm d}\boldsymbol{s}{\rm d}\boldsymbol{u} \frac{\TH^2}{\Omega^2} \frac{{\rm e}^{\frac{{\rm i}}{\hbar}\boldsymbol{s}\boldsymbol{u}}}
{t_{{\rm enc},1}t_{{\rm enc},2}}
\int_{\frac{t_{{\rm encs}}}{2k+1}}^{\frac{t_{{\rm enc,min}}}{k}}
{\rm d}\tau_p\nonumber\\
&&\times{\rm e}^{-p_{2k+1}\tau_p/\tau_D}\nonumber\\&&\times\left[{\rm e}^{-\left(p_{2k+2}-p_{2k+1}\right)
\left(t_{{\rm encs}}-2k\tau_p\right)/\tD}\right.\nonumber\\
&& {} -{\rm e}^{-\left(p_{2k+3}-p_{2k+2}\right)
\left(t_{\rm enc,min}-k\tau_p\right)/\tD} \nonumber\\
&&\left.\times{\rm e}^{-\left(p_{2k+2}-p_{2k+1}\right)
\left(t_{\rm enc,max}-k\tau_p\right)/\tD}\right] ,
\end{eqnarray}
and second for $t_{\rm enc,max}^p<\tau_p<t_{{\rm enc},1}^p+t_{{\rm enc},2}^p$ by
\begin{eqnarray}
\label{ap32}
{\rm var}\,G^{[{\ref{ana4}({\rm a})}]}_{\rm mt(2)}&=&
\frac{2N_{\rm L}^2N_{\rm R}^2}{\left(N_{\rm L}+N_{\rm R}\right)^4}\frac{2\tD}{\left(p_{2k+3}+p_{2k+1}-2p_{2k+2}\right)}\nonumber \\
&&\times
\int_{-c}^c{\rm d}\boldsymbol{s}{\rm d}\boldsymbol{u} \frac{\TH^2}{\Omega^2} \frac{{\rm e}^{\frac{{\rm i}}{\hbar}\boldsymbol{s}\boldsymbol{u}}}
{t_{{\rm enc},1}t_{{\rm enc},2}}
\int_{\frac{t_{{\rm enc,max}}}{k+1}}^{\frac{t_{{\rm encs}}}{2k+1}}
{\rm d}\tau_p \nonumber\\
&&\times\left[{\rm e}^{-\left(
2p_{2k+2}-p_{2k+3}\right)\tau_p/\tD}\right.\nonumber\\
&&\times{\rm e}^{-\left(p_{2k+3}-p_{2k+2}\right)
\left(t_{{\rm encs}}-2k\tau_p\right)/\tD} \nonumber\\
&& {} -{\rm e}^{-\left(p_{2k+3}-p_{2k+2}\right)
\left(t_{\rm enc,min}-k\tau_p\right)/\tD} \nonumber \\
&&\times{\rm e}^{-\left(p_{2k+2}-p_{2k+1}\right)
\left(t_{\rm enc,max}-k\tau_p\right)/\tD} \nonumber \\
&&\left.\times{\rm e}^{-p_{2k+1}\tau_p/\tD}\right].
\end{eqnarray}
In the case of the two encounters overlapping at both ends the contribution is
obtained to be
\begin{eqnarray}
\label{ap4}
{\rm var}\,G^{[{\ref{ana4}({\rm b})}]}_{\rm
mt}&=&\frac{2N_{\rm L}^2N_{\rm R}^2}{\left(N_{\rm L}+N_{\rm R}\right)^4}
\int_{-c}^c{\rm d}\boldsymbol{s}{\rm d}\boldsymbol{u}
\int_{\frac{t_{{\rm enc,max}}}{k+1}}^{\frac{t_{{\rm encs}}}{2k+1}}
{\rm d}\tau_p \frac{\TH^2}{\Omega^2} \nonumber\\
&&\times\frac{{\rm e}^{\frac{{\rm i}}{\hbar}\boldsymbol{s}\boldsymbol{u}}}
{t_{{\rm encs}}}\left[t_{{\rm encs}}-\left(2k+1\right)\tau_p\right]\nonumber\\
&&\times{\rm e}^{-\left(p_{2k+3}-p_{2k+2}\right)
\left(t_{{\rm encs}}-2k\tau_p\right)/\tD}\nonumber\\
&&\times{\rm e}^{-\left(2p_{2k+2}-p_{2k+3}\right)\tau_p/\tD}.
\end{eqnarray}

\section{Different tunneling probabilities} \label{diffprobs}

Here we generalize \Eref{li22.1} to the case of different
$\Gamma_j$ for the different lead modes. Factors like $\Gamma N$ are then
replaced by a sum over the $\Gamma_j$ with respect to the $N$ open channels. An
analogous replacement is made for the $p_j$ for $j\geq1$. In order to keep the
notation compact we introduce
\begin{equation}
\label{ap5}
G_{(i)}\equiv\sum_{j=1}^{N_i}\Gamma_j,\qquad G_i\equiv\sum_{j=1}^{N_{\rm L}+N_{\rm R}}
p_j.
\end{equation}
For example, the contribution from the diagram in Fig.\ \ref{ana1}(a),
previously given in \eref{li0.3}, becomes
\begin{equation}
\label{ap6}
{\rm var}\,G^{[{\ref{ana1}}]} = \frac{G_{(1)}^{2}G_{(2)}^{2}G_{2}^{2}}{G_{1}^{6}}.
\end{equation}
where the Ehrenfest-time dependence is the same as in \eref{li0.3} although we set
$\tE=0$ here.  The first two terms in the numerator result from the channel 
summations. The third term replaces the factor $p_2^2N^2$ in \eref{li0.3} and the 
denominator replaces the previous $(\Gamma N)^4$. 

To keep the contributions from configurations where one link connecting the
encounter to the opening is removed in a compact form, we define
\begin{equation}
\label{ap7}
H_{(i),k}\equiv\sum_{j=1}^{N_i}\Gamma_j\left(1-\Gamma_j\right)^{k-1}.
\end{equation}
Considering again the corresponding contribution from Fig.\
\ref{ana1}(a), previously given in \eref{li0.4}, we have
\begin{equation}
\label{ap8}
{\rm var}\,G^{[{\ref{ana1}-1{\rm l}}]} =
-\frac{2\left[H_{(1),2}G_{(2)}+G_{(1)}H_{(2),2}\right]G_{2
}G_{(1)}G_{(2)}}{G_{1}^{5}}.
\end{equation}
The first term derives from when the link connecting the
encounter to lead 1 is shrunk while the second term corresponds to when the link 
connecting an encounter to lead 2 is removed. Compared to \eref{ap6},
the factor $H_{(i),k}$ takes into account that the particle is entering the 
system in a certain channel and returns to the same channel after
traversing a link.

Removing two links from different encounters likewise leads to
\begin{equation}
\label{ap8.1}
{\rm var}\,G^{[{\ref{ana1}(a)-2{\rm l}}]} =
\frac{4G_{(1)}H_{(1),2}H_{(2),2}G_{(2)}}{G_{1}^{4}}.
\end{equation}
For the diagram in Fig.\ \ref{ana1}(a) we can also remove two links from the same
encounter, and for this we define
\begin{equation}
\label{ap9}
I_{(i),k}\equiv\sum_{j=1}^{N_i}\Gamma_j^2\left(1-\Gamma_j\right)^{k-1} ,
\end{equation}
so that we obtain
\begin{eqnarray}
\label{ap8.2}
{\rm var}\,G^{[{\ref{ana1}(a)-2{\rm l(s)}}]}\!&=&\!
-\frac{\left[I_{(1),1}G_{(2)}^{2}+G_{(1)}^{2}I_{(2),1}\right]G_{2}}{G_{1}^{4}} ,\\
%
%
\label{ap8.3}
{\rm var}\,G^{[{\ref{ana1}(a)-3{\rm l}}]}\!&=&\!
\frac{2\left[G_{(1)}H_{(1),2}I_{(2),1}+I_{(1),1}H_{(2),2}G_{(2)}\right]}{G_{1}^{3}} ,\nonumber\\
\\
%
%
\label{ap8.4}
{\rm var}\,G^{[{\ref{ana1}(a)-4{\rm l}}]}\!&=&\!
\frac{I_{(1),1}I_{(2),1}}{G_{1}^{2}} .
\end{eqnarray}
For the diagram in Fig.\ \ref{ana1}(b) we cannot shrink two links attached to the 
same encounter (or more than two links) and because of the way the encounters are arranged
when we shrink two links we obtain
\begin{eqnarray}
\label{ap8.5}
{\rm var}\,G^{[{\ref{ana1}(b)-2{\rm l}}]} &=&
\frac{2G_{(1)}H_{(1),2}H_{(2),2}G_{(2)}}{G_{1}^{4}} \\
&& {} +\frac{H_{(1),2}^{2}G_{(2)}^{2}+G_{(1)}^{2}H_{(2),2}^{2}}{G_{1}^{4}} . \nonumber
\end{eqnarray}
Using the definitions above, this is equal to \eref{ap8.1} plus the results from
(\ref{ap8.2}--\ref{ap8.4}) so that time reversal symmetry still leads simply 
to a factor of 2.

In an analogous manner, we can show that each diagram in Fig.\ \ref{ana2} gives
twice the contribution of the diagram in Fig.\ \ref{ana1}(b) while for the 
diagrams in Fig.\ \ref{ana3} we obtain
\begin{eqnarray}
\label{ap8.6}
{\rm var}\,G^{[{\ref{ana3}}]} &=& -\frac{2G_{(1)}^{2}G_{(2)}^{2}G_{3}}{G_{1}^{5}},\\
%
%
\label{ap8.7}
{\rm var}\,G^{[{\ref{ana3}-1{\rm l}}]} &=& 
\frac{4\left[G_{(1)}H_{(1),3}G_{(2)}^{2}+G_{(1)}^{2}H_{(2),3}G_{(2)}\right]}
{G_{1}^{4}},\nonumber\\ 
\\
%
%
\label{ap8.8}
{\rm var}\,G^{[{\ref{ana3}-2{\rm l}}]} &=& 
\frac{2\left[I_{(1),2}G_{(2)}^{2}+G_{(1)}^{2}I_{(2),2}\right]}{G_{1}^{3}} .
\end{eqnarray}

Summing all these contributions, we obtain the RMT result \cite{BroRMT} 
for the leading order in $N$ contribution to the conductance variance
\begin{eqnarray} 
\label{condfluctresulteqn}
{\rm var}\,G^{\rm RMT}&=&
\frac{1}{\left(g_{1}+g'_{1}\right)^{6}}\left[2g_{1}^{5}g'_{2}-2g_{1}^{4}
g'_{2}g'_{1}-4g_{1}^{3}g'_{2}{g'_{1}}^{2}\right.\nonumber\\
&&\left.{}-4g_{1}^{2}g_{2}{g'_{1}}^{3}-2g_{1}g_{2}{g'_{1}}^{4}+
2g_{2}{g'_{1}}^{5}+2g_{1}^{4}{g'_{1}}^{2}\right.\nonumber\\
&&\left.{}+4g_{1}^{3}{g'_{1}}^{3}+2g_{1}^{2}{g'_{1}}^{4}
+3g_{1}^{4}{g'_{2}}^{2}+6g_{1}^{2}g_{2}g'_{2}{g'_{1}}^{2}\right.\nonumber\\
&&\left.{}+3g_{2}^{2}{g'_{1}}^{4}-2g_{1}^{5}g'_{3}
-2g_{1}^{4}g'_{3}g'_{1}-4g_{1}g_{3}{g'_{1}}^{4}\right.\nonumber\\
&&\left.{}-2g_{3}{g'_{1}}^{5}\right] ,
\end{eqnarray}
for systems without time reversal symmetry, and twice this result for those with.
Here the notation introduced in \cite{BroRMT}
\begin{equation}
\label{ap10}
g_{k}=\sum_{n=1}^{N_{\rm L}}\Gamma_n^k,\qquad g'_{k}=\sum_{n=1}^{N_{\rm R}}\Gamma_n^k ,
\end{equation}
was used.

\section{Shot noise}

\begin{figure}
\centerline{\includegraphics[width=0.8\columnwidth]{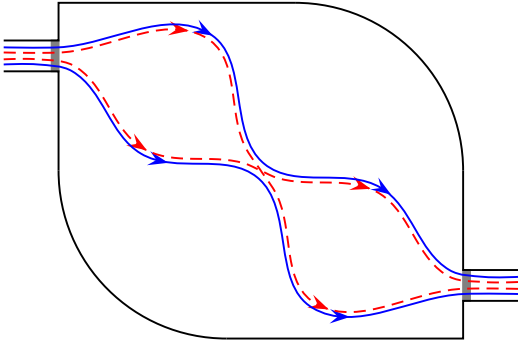}}
\caption{Two trajectories with a single encounter and two partner trajectories.}
\label{trajectoryquad}
\end{figure}

Here we calculate the shot noise power. It can be written as
\cite{Buttik}
\begin{equation}
P=G-h= \left\langle {\rm Tr}\left[\mathbf{tt}^{\dagger}\right]\right\rangle
- \left\langle {\rm Tr}\left[\left(\mathbf{tt}^{\dagger}\right)^2\right]\right\rangle.
\end{equation}
The average conductance $G(E)$ has previously been calculated \cite{Whitney,Kuip},
and the first few terms for systems with time reversal symmetry can be written as
\begin{eqnarray}\label{Gdiffprobeq}
G(E)&=&\frac{G_{(1)}G_{(2)}}{G_1}-\left(\frac{2}{\beta}-1\right)\frac{G_{(1)}G_{(2)}G_{2}}{G_{1}^{3}} \\
&& {} +\left(\frac{2}{\beta}-1\right)\frac{G_{(1)}H_{(2),2}+H_{(1),2}G_{(2)}}{G_{1}^{2}} +\ldots \nonumber ,
\end{eqnarray}
using the notation of Appendix \ref{diffprobs}.  The second two terms derive 
from the diagrams in Fig.\ \ref{fig:orbt} so that without time reversal 
symmetry only the first term in \eref{Gdiffprobeq} remains.  This is included
as $\beta=2$ for systems without time reversal symmetry and $\beta=1$ for
those with time reversal symmetry.
The function $h$ is given semiclassically in terms of 4 trajectories 
\begin{eqnarray} 
\label{shotnoisetraj}
h&=&\left\langle\frac{1}{\TH^{2}}\sum_{\substack{{a,b} \cr {c,d}}}
\sum_{\substack{\gamma (a\to b)\cr \gamma' (c\to b)}}
\sum_{\substack{\xi (c\to d) \cr \xi' (a\to d)}}
A_{\gamma}A_{\gamma'}^{*}A_{\xi}A_{\xi'}^{*}\right.\nonumber\\
&&\left.\hspace{3em}\times {\rm e}^{\frac{{\rm i}}{\hbar}
(S_{\gamma}-S_{\gamma'}+S_{\xi}-S_{\xi'})}\right\rangle,
\end{eqnarray}
where the main difference from the conductance variance \eref{condvartraj} is
that the trajectories connect different channels and the sum is unrestricted. 
We start with the diagram from Fig.\ \ref{trajectoryquad}, where since the partner 
trajectories cross over in the encounter they automatically travel from and to
the correct channels and all four channels are unrestricted.  This means that
this structure is now lower order (in inverse channel number) than it was for
the conductance variance (it was there contained by taking into account the
diagram in Fig.\ \ref{ana1}(a) and removing two links connected to the same encounter).
For the diagram in Fig.\ \ref{trajectoryquad}, we therefore have
\begin{eqnarray} 
\label{xcontsneqn}
h^{[{\ref{trajectoryquad}}]}&=&-\frac{G_{(1)}^{2}{G_{(2)}}^{2}G_2}{G_{1}^4} \nonumber \\
&& {} +\frac{2\left[G_{(1)}H_{(1),2}G_{(2)}^2+G_{(1)}^2H_{(2),2}G_{(2)}\right]}{G_{1}^3} \nonumber \\
&& {} + \frac{G_{(1)}^{2}I_{(2),1}+I_{(1),1}G_{(2)}^{2}}{G_{1}^{2}}.
\end{eqnarray}
This contribution, along with its Ehrenfest-time dependence, was previously
calculated in \cite{Whitney}.

Moving to the next order term, which has recently been calculated using RMT
\cite{rbm08}, we again simply need to look at the corresponding structures and consider
the possible ways of shrinking links.  Without time reversal symmetry there are
no possible structures at this order, so we therefore consider systems with
time reversal symmetry. The structures which contribute are depicted in Fig.\ 2
of \cite{braunetal06}.  There is also, in Fig.\ 2(d) there, a quadruplet involving two
independent pairs, one of which is simply a diagonal pair, while the other
involves a single 2-encounter, as in Fig.\ \ref{fig:orbt} of this paper.  
Because the start (or end) channels must coincide $a=c$ (or $b=d$) 
there is a subtlety when we consider shrinking one link.  
If we shrink the link on the left (or right) side of the 2-encounter so
that it moves into the lead where the diagonal pair emanate (or terminate) we
actually have the same case as when the 3-encounter from Fig.\ 2(c) (of
\cite{braunetal06}) moves into the lead.  As such, we include this case there,
leaving a contribution of
\begin{eqnarray}
\label{dcontsneqn}
h^{[{2(\mathrm{d})}]}&=&-\frac{2\left[G_{(1)}^{2}I_{(2),1}+I_{(1),1}G_{(2)}^{2}\right]G_{2}}
{G_{1}^{4}} \\
&& {} +\frac{2\left[G_{(1)}H_{(1),2}I_{(2),1}+I_{(1),1}H_{(2),2}G_{(2)}\right]}
{G_{1}^{3}}. \nonumber
\end{eqnarray}
The diagram in Fig.\ 2(a) of \cite{braunetal06} is similar to the diagram
in Fig.\ \ref{ana1}(b) here and we obtain
\begin{eqnarray}
\label{acontsneqn}
h^{[{2(\mathrm{a})}]}&=&\frac{4G_{(1)}^{2}G_{(2)}^{2}G_{2}^{2}}{G_{1}^{6}} 
+ \frac{8 G_{(1)}H_{(1),2}H_{(2),2}G_{(2)}}{G_{1}^{4}} \nonumber \\
&& {} -\frac{8\left[G_{(1)}H_{(1),2}G_{(2)}^{2}+G_{(1)}^{2}H_{(2),2}G_{(2)}\right]G_{2}}
{G_{1}^{5}} \nonumber \\
&& {} +\frac{2\left[G_{(1)}^{2}H_{(2),2}^{2}+H_{(1),2}^{2}G_{(2)}^{2}\right]}
{G_{1}^{4}} . 
\end{eqnarray}
For the diagram in Fig.\ 2(b) of \cite{braunetal06} we obtain twice the result of
Fig.\ \ref{ana2}(a) here [or four times the result of Fig.\ \ref{ana1}(b)] for the 
conductance variance.  For the diagram in Fig.\ 2(c) of \cite{braunetal06} we likewise
obtain twice the result of Fig.\ \ref{ana3}(a) here for the conductance variance. 
With all the next to leading order shot noise contributions, we can combine 
them and indeed find the same result as in \cite{rbm08}. 

We can simplify the result by setting all of the tunneling probabilities
equal to $\Gamma$:
\begin{eqnarray}
P(\Gamma)&=&\frac{\Gamma(1-\Gamma)N_{1}N_{2}}{\left(N_{\rm L}+N_{\rm R}\right)}
+\frac{\Gamma(3\Gamma-2)N_{1}^{2}N_{2}^{2}}{\left(N_{\rm L}+N_{\rm R}\right)^{3}} \nonumber \\
&& {} +\left(\frac{2}{\beta}-1\right)
\left(\frac{\Gamma(4\Gamma-3)N_{1}N_{2}\left(N_{1}-N_{2}\right)^{2}}
{\left(N_{\rm L}+N_{\rm R}\right)^{4}}\right) \nonumber \\
&& {} +\ldots . 
\end{eqnarray}
Setting $\Gamma=1$, we also recreate the first two terms, 
\begin{eqnarray}
P(\Gamma=1)&=&\frac{N_{1}^{2}N_{2}^{2}}{\left(N_{\rm L}+N_{\rm R}\right)^{3}} \\
&& {} +\left(\frac{2}{\beta}-1\right)\left(\frac{N_{1}N_{2}\left(N_{1}-N_{2}\right)^{2}}
{\left(N_{\rm L}+N_{\rm R}\right)^{4}}\right)+\ldots, \nonumber
\end{eqnarray}
of the result in \cite{braunetal06}.

\end{document}